\documentclass[fleqn,usenatbib]{mnras}

\usepackage{newtxtext,newtxmath}

\usepackage[T1]{fontenc}

\usepackage{graphicx}	
\usepackage{amsmath}	
\usepackage[normalem]{ulem}

\usepackage{upgreek,xspace}

\newcommand{\mum}{\ensuremath{\upmu\mathrm{m}}\xspace}

\title[Three's a crowd]{Keplerian motion of a compact source orbiting the inner disc of PDS 70: a third protoplanet in resonance with b and c?}

\author[I. Hammond et al.]{Iain Hammond$^{1,\,2}$\thanks{E-mail: dr.iain.hammond@gmail.com},
Valentin Christiaens$^{3,\,4}$,
Daniel J. Price$^{1}$, Dori Blakely$^{5,\,6}$, David Trevascus$^{2}$,
\newauthor{Markus J. Bonse$^{7,\,8}$, Faustine Cantalloube$^{9}$, Gabriel-Dominique Marleau$^{2,\,10,\,11}$, Christophe Pinte$^{1,\,9}$,}
\newauthor{Sandrine Juillard$^{3}$, Matthias Samland$^{2}$, William Thompson$^{6}$ and Alex Wallace$^{1}$}\\
$^{1}$ School of Physics \& Astronomy, Monash University, Vic 3800, Australia\\
$^{2}$ Max-Planck-Institut f\"ur Astronomie, K\"onigstuhl 17, 69117 Heidelberg, Germany\\
$^{3}$ Space sciences, Technologies \& Astrophysics Research (STAR) Institute, Universit\'e de Li\`ege, All\'ee du Six Ao\^ut 19c, B-4000 Sart Tilman, Belgium\\
$^{4}$ Institute of Astronomy, KU Leuven, Celestijnenlaan 200D, Leuven,
Belgium\\
$^{5}$ Department of Physics and Astronomy, University of Victoria, 3800 Finnerty Road, Elliot Building, Victoria, BC V8P 5C2, Canada\\
$^{6}$ NRC Herzberg Astronomy and Astrophysics, 5071 West Saanich Road, Victoria, BC V9E 2E7, Canada\\
$^{7}$ ETH Zurich, Institute for Particle Physics \& Astrophysics, Wolfgang-Pauli-Str. 27, 8093 Zurich, Switzerland \\
$^{8}$ Max-Planck-Institut f\"ur Intelligente Systeme, Max-Planck-Ring 4, 72076 T\"ubingen, Germany\\
$^{9}$ Universit\'e Grenoble Alpes, CNRS, IPAG, F-38000 Grenoble, France\\
$^{10}$ Fakult\"at f\"ur Physik, Universit\"at Duisburg--Essen,
Lotharstra\ss{}e 1,
47057 Duisburg, Germany\\
$^{11}$ Physikalisches Institut,
Universit\"{a}t Bern,
Gesellschaftsstr.~6,
3012 Bern, Switzerland}
\date{Accepted 2025 April 4. Received 2025 April 3; in original form 2024 December 13}

\pubyear{2024}

\begin{document}
\label{firstpage}
\pagerange{\pageref{firstpage}--\pageref{lastpage}}
\maketitle
\begin{abstract}
The disc around PDS 70 hosts two directly imaged protoplanets in a gap. Previous VLT/SPHERE and recent \textit{James Webb Space Telescope}/NIRCam observations have hinted at the presence of a third compact source in the same gap at $\sim$13~au, interior to the orbit of PDS~70~b.
We reduce seven published and one unpublished VLT/SPHERE datasets in \textit{YJH} and \textit{K} bands, as well as an archival VLT/NaCo dataset in \textit{L'} band, and an archival VLT/SINFONI dataset in \textit{H+K} band. We combine angular-, spectral- and reference star differential imaging to search for protoplanet candidates.
We recover the compact source in all epochs, consistent with the JWST detection, moving on an arc that can be fit by Keplerian motion of a protoplanet which could be in a resonance with PDS~70~b \& c. We find that the spectral slope is overall consistent with the unresolved star and inner disc emission at 0.95--1.65~\mum, which suggests a dust scattering dominated spectrum. An excess beyond 2.3~\mum could be thermal emission from either a protoplanet or heated circumplanetary dust, variability, or inner disc contamination, and requires confirmation. While we currently cannot rule out a moving inner disc feature or a dust clump associated with an unseen planet, the data supports the hypothesis of a third protoplanet in this remarkable system.

\end{abstract}




\section{Introduction}\label{sec:intro}
Mean-motion resonance (MMR) capture is expected to be a key mechanism that sets the orbital architecture of protoplanets during the disc phase and halts migration towards the stellar potential well \citep{Weiss:2023,Krijt:2023}. MMR is observed in the Galilean satellites Io, Europa and Ganymede around Jupiter \citep{Laplace1799} and would explain the coplanar 4:2:1 resonance of the Gliese-876 exoplanets \citep{Nelson:2016, Cimerman:2018}, of which two of the planets are several times the mass of Jupiter. Imaging of the young ($\sim$30--60 Myr) system HR~8799 revealed four giant planets consistent with a coplanar 8:4:2:1 Laplace resonance for long-term stability \citep{Marois:2008tk, Marois:2010ue,Fabrycky:2010,Gozdziewski:2020}. However, our understanding of how and when protoplanets enter such a configuration is limited due to the small number of confirmed multi-protoplanet systems.

PDS~70 (V1032 Centauri) is a $\sim$5Myr, 0.87 -- 1$M_\odot$ T~Tauri star \citep{Hashimoto:2012wi, Muller:2018wg, Long:2018, Wang:2021a} at a distance of 112.39$\pm$0.24~pc \citep{Gaia-Collaboration:2023} surrounded by a protoplanetary disc. The system hosts two directly imaged protoplanets (hereafter PDS 70 b \& c) with infrared emission, H$\alpha$ emission, and a sub-mm continuum counterpart \citep{Keppler:2018vt, Muller:2018wg, Haffert:2019td, Benisty:2021uj}. PDS 70 b \& c have a deprojected separation of $\sim$20 and 34~au from the star, with orbital fits of b preferring non-zero eccentricity \citep{Wang:2021a}. Dedicated hydrodynamical simulations found that the protoplanets lock into a 2:1 MMR \citep{Bae:2019}, and migrate to the same configuration even if initially placed on wider orbits \citep{Toci:2020wu}.

\citet{Mesa:2019a} first identified a candidate third protoplanet interior to the orbit of b at $\sim$13~au with the Spectro-Polarimetic High contrast imager for Exo-planets REsearch \citep[SPHERE;][]{Beuzit:2008} intrument at the VLT. Their ``point-like feature'' (PLF) had a spectrum similar to the stellar spectrum between 0.95 -- 1.65~\mum and it was marginally detected in \textit{H} and \textit{K}-bands. Given the uncertain radial extend of the inner disc around PDS~70~A, they concluded the PLF is likely forward scattering from disc material, although an embedded planet was not ruled out. Recent JWST/NIRCam observations with the F187N (1.87~\mum) filter in \citet{Christiaens:2024} reported a bright signal near the previously reported PLF and was robust to the subtraction of an optimal disc model. The measured astrometry is consistent with Keplerian motion of an object at $\sim$13.5~au in the plane of the disc. The feature was interpreted as either tracing a dusty circumplanetary envelope or disc, or a dust clump, and tentatively labelled PDS~70~d.

In this paper, we revisit archival data sets using novel and state-of-the-art high-contrast post-processing methods to constrain the nature and orbit of candidate protoplanet PDS~70~d. We present the high-contrast imaging data of PDS~70 and data reduction in Section \ref{sec:observations}. In  Section~\ref{sec:results}, we present astrometry, orbital fitting and the spectrum of candidate d. We discuss the potential planetary nature of the source in Section~\ref{sec:discussion} and present our conclusions in Section~\ref{sec:conclusion}.

\begin{table*}
\begin{center}
\caption{Summary of high-contrast imaging observations of PDS~70 reduced and analyzed in this work. FITS files are in the Data Availability statement.}
\label{tab:obs}
\begin{tabular}{lccccccccccccc}
\hline
\hline
Date & Strategy & Program ID & Instrument & Filter/ & Plate scale$^{\rm (a)}$ & Coronagraph & DIT & 
T$_{\rm int}^{\rm (b)}$ & $\langle\epsilon\rangle^{\rm (c)}$ & $\Delta$PA$^{\rm (d)}$ \\
& &  & & Prism & [mas px$^{-1}$] & & [s] & 
[min] & [\arcsec] & [\degr] \\
\hline
2012 Mar 31$^{\rm (e)}$ & ADI & GS-2012A-C-3 & NICI & \emph{L'} & 17.95$\pm 0.01$  & -- & 0.76 & 33.57 & -- & 99.37 \\

2014 May 10 & SADI & 093.C-0526 & SINFONI & \textit{K} & 12.5 & -- & 1 & 116 & 1.13 & 99.8 \\

2016 June 01 & ADI & 097.C-0206(A) & NaCo & \emph{L'} & 27.193$\pm 0.09$  & -- & 0.2 & 62 & 0.5 & 85 \\

2018 Feb 25 & ADI & 1100.C-0481(D) & IRDIS & \emph{K$_1$K$_2$} & 12.63$\pm 0.009$  & N\_ALC\_YJH\_S & 96 & 112 & 0.4 & 91.86 \\

2018 Feb 25 & ASDI & 1100.C-0481(D) & IFS & \emph{YJH} & 7.46$\pm 0.02$  & N\_ALC\_YJH\_S & 96 & 136 & 0.4 & 91.86 \\

2019 Mar 06 & ASDI & 1100.C-0481(L) & IFS & \emph{YJH} & 7.46$\pm 0.02$ & N\_ALC\_Ks & 96  & 76.8 & 0.39 & 56.4 \\

2021 May 17 & ASDI & 1104.C-0416(E) & IFS & \emph{YJH} & 7.46$\pm 0.02$ & N\_ALC\_YJH\_S & 96 & 76.8 & 0.46 & 56.38 \\

2021 Aug 21 & ASDI/RDI & 107.22UJ.001 & IFS & \emph{YJH} & 7.46$\pm 0.02$ & N\_ALC\_YJH\_S & 32 & 27.2 & 0.75 & 14 \\


2021 Sep 03 & ASDI/RDI & 107.22UJ.001 & IFS & \emph{YJH} & 7.46$\pm 0.02$ & N\_ALC\_YJH\_S & 32 & 21.3 & 0.54 & 5.1 \\

2022 Feb 28 & ADI/RDI & 107.22UJ.001 & IRDIS & \emph{K\textsubscript{s}} & 12.265$\pm 0.009$ & N\_ALC\_YJH\_S & 16 & 36.2 & 0.48 & 19.51 \\

2022 Feb 28 & ASDI/RDI & 107.22UJ.001 & IFS & \emph{YJH} & 7.46$\pm 0.02$ & N\_ALC\_YJH\_S & 32 & 36.2 & 0.48 & 19.91 \\

2023 Mar 8 & Roll & 1282 & NIRCam & F187N & 30.63$\pm 0.06$ & -- & 0.35 & 8.3 & -- & 5.0 \\

\hline
\end{tabular}
\end{center}
Notes: $^{\rm (a)}$\citet{Maire:2016vq} for IRDIS \& IFS, \citet{Launhardt:2020wu} for NaCo, \citet{Keppler:2018vt} for NICI. $^{\rm (b)}$Total integration time on PDS~70 
after bad frame removal. $^{\rm (c)}$Average seeing at $\lambdaup$~=~500nm over all PDS~70 integrations. 
$^{\rm (d)}$Parallactic angle variation after bad frame removal. $^{\rm (e)}$Appendix \ref{sec:nici_data}.
\end{table*}

\section{Observations and Data Reduction}\label{sec:observations}

\subsection{VLT/SPHERE IRDIFS observations}\label{sec:sphere_data}
We processed six epochs of coronagraphic observations taken with the SPHERE instrument in IRDIFS mode --- that is, simultaneous acquisition with the Infrared Differential Imaging Spectrometer (IRDIS) and the Integral Field Spectrograph (IFS) sub-instruments. These are summarised in Table \ref{tab:obs}. We re-analysed five observations that were presented in \citet{Mesa:2019a} and \citet{Wahhaj:2024}. The 2021 May 17 epoch is unpublished data from the SPHERE infrared survey for exoplanets (SHINE, \citealt{Desidera:2021}). Each sequence was taken in pupil stabilised mode, enabling the use of angular differential imaging (ADI; \citealt{Marois:2006vd}) with IRDIS and the IFS and spectral differential imaging (SDI; \citealt{Sparks:2002}) with the IFS. Three epochs from ESO program 107.22UJ.001 (PI: Z. Wahhaj, \citealt{Wahhaj:2024}) observed a nearby PSF reference star, UCAC2 14412811, interspersed between every 10 minutes of integration time on PDS~70, enabling `star-hopping' \citep{Wahhaj:2021} reference-star differential imaging (RDI).

\begin{figure*}
    \includegraphics[width=0.98\textwidth]{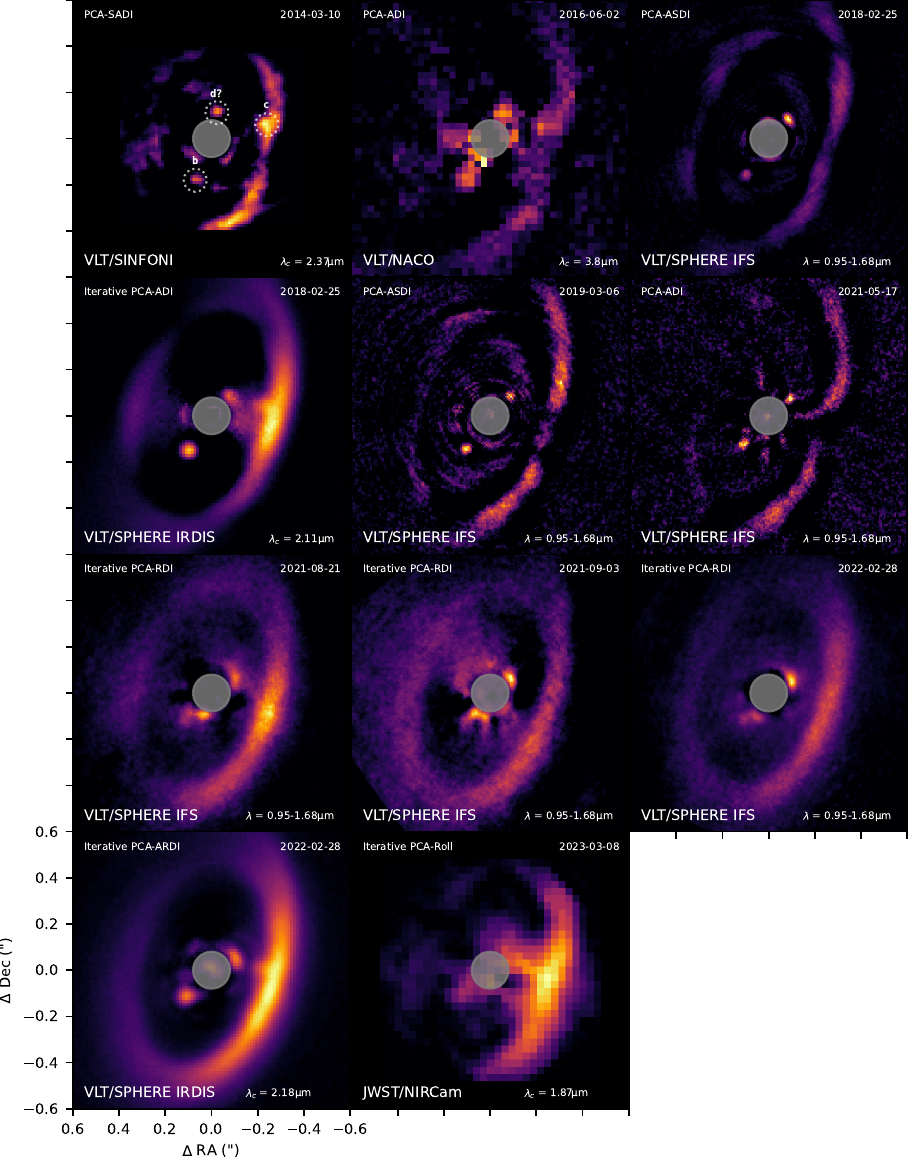}
    \caption{Nine years of high-contrast imaging data of PDS~70. All images are cropped to a 1\farcs2 $\times$ 1\farcs2 field of view centred on the star and in chronological order (left to right, top to bottom). The semi-transparent mask over A is 160~mas wide, roughly the inner working angle of the SPHERE coronagraph (SINFONI, NaCo and NIRCam are non-coronagraphic). The PSF-subtraction strategy is indicated in the top-left corner of the frame. The disc, b and candidate d are detected in all frames, whereas c is clearest at $\lambda>2~\mum$. North is up, East is to the left. Figure made using \textsc{casa cube} (\url{https://github.com/cpinte/casa_cube}).
    }
    \label{fig:observations}
\end{figure*}

All SPHERE data were reduced with an in-house open source pipeline, {\sc vcal-sphere}\footnote{\url{https://github.com/VChristiaens/vcal_sphere}} \citep{Christiaens:2023a}, which uses \textsc{esorex} (\textsc{v3.13.8}) recipes for calibration and Vortex Image Processing (\textsc{vip}; \citealt{Gomez-Gonzalez:2017uw,Christiaens:2023}) functions for pre- and post-processing. 

For IRDIS, the pipeline involved the same steps as presented in \cite{Hammond:2023}; (i)~flat-fielding, (ii)~principal component analysis (PCA)-based sky subtraction \citep{Hunziker:2018ux, Ren:2023a}, (iii)~bad pixel correction, (iv)~centering of the coronagraphic images via satellite spots, (v)~centering of the non-coronagraphic unsaturated PSF images using 2D Gaussian fits, (vi)~anamorphism correction, (vii)~bad-frame removal, (viii)~final PSF creation, and (ix)~post-processing with median-angular differential imaging \citep[ADI, ][]{Marois:2006tx}, PCA-ADI in full-frame \citep{Amara:2012um,Soummer:2012wx}, annular PCA-ADI \citep{Absil:2013}, and, when relevant, PCA-RDI. Each half of the detector was reduced independently for data sets taken in dual band imaging (DBI) mode. As the 2018 Feb 25 ADI and 2022 Feb 28 RDI data sets had the longest integration time on target and excellent observing conditions, we consider only these two datasets to extract the photometry of candidate~d in the $K_1$$K_2$ and $K_s$ filters, respectively. The location of d near the inner working angle (IWA) of the coronagraph makes photometry extraction unreliable in datasets acquired in more challenging conditions.

Images obtained with PCA-ADI are affected by both self- and over-subtraction which cause geometric artefacts to extended signals (see \citealt{Juillard:2022} for the expected effect on the disc of PDS~70), while PCA-RDI reductions are still affected by over-subtraction. To alleviate these geometric artefacts, we also post-processed the 2018 Feb 25 and 2022 Feb 28 data sets using iterative PCA \citep[IPCA;][]{Pairet:2021}-ADI and IPCA-ARDI, respectively \citep{Juillard:2023, Juillard:2024}. Given the stability of the 2018 dataset and the close separation of d, the number of principal components ($n_{\rm pc}$) was set to 1 and the number of iterations ($n_{\rm it}$) to 10 for the IPCA-ADI reduction. For the 2022 dataset, the science and reference stellar residual signals were not well correlated near the edge of the coronagraphic mask, hence we opted for an ARDI strategy with more principal components. Signal from d is detected with PCA-ARDI for $n_{\rm pc}=2$--$16$, with an optimal signal-to-noise ratio (SNR) achieved with 7 to 10 components. This reduction leads to significant self- and over-subtraction of the outer disc. For a faster convergence of IPCA-ARDI, we initialized its first iteration with the result of PCA-RDI obtained with data imputation \citep{Ren:2023b} instead of PCA-ARDI. As opposed to \citet{Ren:2023b}, we set the anchor mask to lie inside the cavity because in the case of PDS~70, disc signals extend up to the control ring. We selected an empty patch of signal in the cavity covering a wedge from PA=-30$\degr$ to PA=15$\degr$~from $0\farcs2$ to $0\farcs3$ separation. 
We selected $n_{\rm pc}=7$ for the IPCA-ARDI reduction, and the algorithm converged to a final image after $n_{\rm it} = 126$ iterations. Finally, we also tested the {\sc mayonnaise}\footnote{\url{https://github.com/bpairet/mayo_hci}} algorithm \citep{Pairet:2021} on the 2018 Feb 25 ADI dataset, which attempts to separate point sources from extended circumstellar signals. The resulting image is shown in Fig.~\ref{fig:andromeda+mago_fig}.

Calibration of IFS data included additional steps to the above. Master detector and instrument flats, spectra positions, and wavelength calibration were computed using \textsc{esorex} to produce calibrated spectral cubes of 39 channels. \textsc{vip} handled pre-processing in a similar fashion as for IRDIS. We leveraged the radial expansion of the satellite spots by $\sim$14$\lambda$/D to determine the scaling factors used for spectral differential imaging, and to identify channels with the highest SNR for the purposes of bad frame detection. 
We exploit both angular and spectral differential imaging using PCA for improved subtraction of stellar speckles. We performed PCA in one step (i.e., PCA-ASDI) in order to reach the highest possible contrast in the absence of a dedicated reference star observation \citep[e.g.,][]{Christiaens:2021, Kiefer:2021}. A rotation threshold of 0.5$\times$FWHM at the separation of candidate~d on each side of the considered frame was used to build the reference PSF library. We explore $n_{\rm pc}=1-20$ and the frames in Fig.~\ref{fig:observations} are shown at $n_{\rm pc}=10$, however we found that the speckles were sufficiently well subtracted after a few $n_{\rm pc}$. Since images obtained with PCA-ASDI are also prone to geometrical artefacts \citep{Christiaens:2019}, we post-processed all datasets acquired in star-hopping mode with IPCA-RDI, applied channel per channel. We then median-combined the residual images, either band-wise (Y, J and H) or over all spectral channels. For the dataset acquired in the best conditions (2022 Feb 28), we also show the IPCA-RDI results obtained channel per channel in Figure~\ref{fig:all_channels}. For two out of the three datasets, the choice of $n_{\rm pc}$ did not matter, as all tested values between 1 and 8 all led to very similar final images. For the 2021 Aug 21 dataset, a minimum $n_{\rm pc}$ value of 5 is required for a visual detection of candidate~d, as the fluctuating observing conditions required a higher $n_{\rm pc}$ value to capture most of the variance in the data. Most over-subtraction effects are then corrected within a few iterations. All IFS results obtained with IPCA-RDI are shown for $n_{\rm pc} = 7$ and $n_{\rm it} = 5$ in Fig.~\ref{fig:observations}.

Finally, we re-analysed two SPHERE datasets from 2015 May 3 and 2015 May 31. These data were taken in poor conditions and with the \textit{YJ} prism instead of the wider \textit{YJH} prism. The narrower spectral range reduces the effectiveness of SDI by excluding the \textit{H}-band. Since not even planet b was detected, we did not proceed with further analysis. As an additional test we processed both 2015 epochs with a second, independent pipeline, \textsc{charis-dep} \citep{Samland:2022}, with much the same result.

\subsection{VLT/NaCo L'-band angular differential imaging observations}\label{sec:naco_data}
We made use of published non-coronagraphic VLT/NaCo \textit{L'}-band (3.8~\mum) observations from the ISPY survey \citep{Launhardt:2020wu} first presented in \citet{Keppler:2018vt}. These data were taken in pupil stabilised mode enabling the use of angular differential imaging. The data were acquired without a coronagraph, using a three-point dither pattern to subtract the thermal background. The sequence has a total integration time of 62 minutes on PDS~70 in good seeing (0\farcs5). We use {\sc PynPoint}\footnote{\url{https://github.com/PynPoint/PynPoint}} \citep{Stolker:2019} for the calibration and pre-processing of the data. Our pre-processing consists of: (i) basic dark and flat calibrations, (ii) a simple thermal background subtraction using sky frames from other dither positions, (iii) a bad pixel interpolation, (iv) an alignment and centering of the star based on cross-correlations, and (v) a PCA-based bad frame rejection routine. After these pre-processing steps, we mean combined every five frames, resulting in the final cube of 1615 frames with 85$\degr$ of parallatic angle variation.
We used annular PCA-ADI in a 2\farcs7 subframe centred on the star and a rotation threshold of 0.5$\times$FWHM at the separation of PDS~70~d and explored $n_{\rm pc}$ = 1--50 with $n_{\rm pc}$ = 5 shown in Fig.~\ref{fig:observations}.

\subsection{VLT/SINFONI K-band spectral differential imaging observations}\label{sec:sinfoni_data}
We revisited VLT/SINFONI data from 2014 May 10 taken with the \textit{H} + \textit{K} grating. The total integration time was 116 minutes (116 spectral cubes of 60s integration each) which resulted in 99\fdg8 of parallactic angle variation in average observing conditions. Due to the 0\farcs8 $\times$ 0\farcs8 field of view, a four-point dithering pattern was used where the star was placed in a different corner of the detector. Data reduction followed a similar approach as detailed in \citet{Christiaens:2019}. Nonetheless, compared to that work, we not only mean-combined quadruplets of spectral cubes obtained at different dither positions to enlarge the field of view of SINFONI, but also produced a different 4D (spectral+temporal) master cube considering a cropped field of view of 0\farcs36 $\times$ 0\farcs36. The latter was obtained by considering all individual spectral cubes where the star was at least 0\farcs18 away from the edge of the field. Since the star drifted on the detector over the course of the observation, only 62 cubes complied to that condition. For both the dither-combined or the cropped master cubes, the 13\% least correlated cubes from the median spectral cube, corresponding to acquisition in the poorest conditions, were subsequently discarded. We then used the {\sc vip} implementation of PCA-SADI to post-process the two different 4D data cubes, that is PCA-SDI on each individual spectral cube of $\sim$ 2000 channels, followed by PCA-ADI on the cube of either 25 or 54 images resulting from PCA-SDI. The number of principal components was set to $n_{\rm pc, S}=1$ or 2 for PCA-SDI, and $n_{\rm pc, A}=1$-$10$ was explored for PCA-ADI. Compared to \citet{Christiaens:2019}, a smaller numerical mask was used to probe separations down to $\sim$62 mas (5 spaxels). While the full $H$+$K$ cube of spectral images was used as PCA library for PCA-SDI, residual images after PCA model subtraction were only mean-combined at wavelengths longer than 2.3~\mum in the $K$ band, where point-like signals benefit from the largest Strehl ratio. Since circumstellar signals are recovered for a range of $n_{\rm pc, A}$ values, while the structure of residual speckle noise varies with $n_{\rm pc, A}$, we median-combine the results obtained with $n_{\rm pc, A}$ ranging from 1 to 8 and from 4 to 10 for the cropped and dither-combined cubes, respectively. Finally we combined into a single image, shown in Fig.~\ref{fig:observations}, the results obtained with PCA-SADI on the cropped and dither-combined cubes in the inner and outer $\sim$0\farcs15 radius, respectively.


\section{Results}\label{sec:results}

\subsection{Moving feature in a gap}\label{sec:moving_source}
Figure \ref{fig:observations} shows our reductions of all data sets including the new 2021 May 17 SPHERE/IFS data set, and JWST/NIRCam data presented in \citet{Christiaens:2024}. We re-detect PDS~70~b in all data sets, whereas PDS~70~c is best detected at $\lambda>2~\mum$ due to the shape of its SED \citep[e.g.,][]{Wang:2020, Christiaens:2024}. The `Point-Like Feature' or `PDS~70~d' is also re-detected in all observations at an average separation of $\sim$110 mas to the north-west of the star.

The source features an extended structure in observations using the RDI strategy. Given the point-like nature of planet b with the RDI strategy, the resolved emission may be genuine and trace authentic circumstellar signals in the direct vicinity of candidate~d (mixed with some contamination from inner disc signal for the case of JWST). When performing only PCA-ADI on the 2022 Feb 28 star-hopping epoch (i.e., not using the reference star), we retrieved a more point-like morphology for d in \textit{YJH} consistent with the earlier SPHERE epochs. However, this observation featured significantly less parallactic angle variation, exacerbating the self-subtraction of extended signals. Without a star-hopping epoch prior to 2021 it is unclear whether this is the true long-term morphology of the `point-like feature' and its compact appearance was simply due to the geometric biases of ADI \citep{Milli:2012ww}. We do, however, notice a dependence on the amount of parallactic angle variation (i.e., 2018 Feb 25, where it is slightly extended compared to shorter sequences).

In order to retrieve the astrometry of candidate d in the SPHERE data, we used the Negative Fake Companion (NEGFC) injection technique \citep{Lagrange:2010} as implemented in \textsc{vip} \citep[described in Appendix D of][]{Christiaens:2021}. First, a Nelder--Mead simplex algorithm was used to obtain a guess on the negative planet flux and its location using $\chi^2_r = \sum\frac{(I- \mu)^2}{\sigma^2}$, where \textit{I} are the pixel intensities measured in a 1.5 $\times$ FWHM aperture at the approximate location of candidate d, and $\mu$ and $\sigma$ the mean and standard deviation of pixel intensities in a 3$\times$FWHM-wide annulus at the separation of d. This estimate was provided to a Markov Chain Monte Carlo (MCMC) algorithm to sample the probability distribution of the (negative) flux, separation and PA, where $\sigma$ was used for scaling the Gaussian log likelihood expression in the MCMC. To check for convergence we used a test based on the integrated autocorrelation time $\tau$ as recommended in the \textsc{emcee} documentation, where the number of samples \textit{N} must be significantly greater than $\tau$. We used \textit{N}/max($\tau$) $>$ 50 for each parameter leading to around 2,500 iterations per data set. We then considered a 30\% burn-in on the MCMC chains. We accounted for the mean radial transmission of the apodised Lyot-stop coronagraph at the location of the injected negative companions (e.g., $\sim$30\% attenuation for a candidate at 110 mas as in the SPHERE manual). We used PCA-RDI for the star-hopping epochs and annular PCA-ADI otherwise. The SPHERE/IFS cubes were median combined in the spectral dimension to improve the SNR, and we considered a 500 $\times$ 500 mas subframe centred on the star (for computational purposes) to build the PSF library. The relative astrometry obtained by the MCMC-NEGFC injection is given in Table \ref{tab:astrometry}. Our final reported astrometric errors are the quadratic sum of a Gaussian fit to the posterior distribution of separation and PA \citep{2017A&A...598A..83W}, true North uncertainty (and 0\fdg1 instrument offset uncertainty for the IFS), plate scale uncertainty, and centering uncertainty of 5 mas for coronagraphic sequences.

As the NaCo data set contained residual signal at the same separation as d in the post-processed images, we opted to use a different technique than for SPHERE to better account for the biases of this noise and lack of pixels to estimate the noise. We used a Nelder-Mead simplex-NEGFC to identify the negative flux that best fits the curvature of the pixel intensities around d, by minimizing the absolute value of the determinant of the Hessian matrix in a 3 $\times$ 3 pixel subframe \citep[][]{Quanz:2015tu, Christiaens:2024}. We removed planet b from the pre-processed cube using the negative flux also found using this technique, and provided the subtracted cube to the \texttt{speckle\_noise\_uncertainty} function of \textsc{vip}. We used this function to inject fake companions at the same separation and flux as estimated for d, but at all azimuths in increments of 1\degr~excluding 10\degr~on either side of d, to estimate their retrieved astrometry. This approach provides a more reliable estimate of the uncertainties associated to residual speckle noise based on the distribution of differences between the injected flux and the retrieved flux for a planet at the same on-sky separation as d. Our final reported astrometric errors in Table \ref{tab:astrometry} for this instrument are the quadratic sum of a Gaussian fit to the distribution of deviations on separation and PA from the injected values, true North uncertainty, and plate scale uncertainty.

\begin{table}
\begin{center}
\caption{Relative astrometry of the candidate PDS~70~d around PDS~70~A obtained with the NEGFC technique described in Section \ref{sec:moving_source}.}
\label{tab:astrometry}
\begin{tabular}{ccccc}
\hline
\hline
Date &Separation (mas)&$\sigma_{\rm sep}$ (mas)&PA (\degr)&$\sigma_{\rm PA}$ (\degr)\\
\hline
2014-05-10 & 117.2 & 12.5 & 348.7 & 4.4 \\
2016-06-02 & 114.3 & 17.1 & 334.6 & 11.7 \\
2018-02-25 & 117.9 & 6.0 & 317.7 & 2.7 \\
2019-03-06 & 120.4 & 8.3 & 315.7 & 2.9 \\
2021-05-17 & 102.4 & 12.1 & 302.6 & 5.6 \\
2021-08-21 & 110.7 & 12.7 & 305.7 & 6.5 \\
2021-09-03 & 106.4 & 10.7 & 299.7 & 6.2 \\
2022-02-28 & 109.3 & 6.5 & 299.4 & 3.9 \\
2023-03-08 & 103.4 & 23.2 & 293.0 & 12.7 \\
\hline
\end{tabular}
\end{center}
Notes: Final errors are a Gaussian fit to the distributions for each parameter, added quadratically with the centering uncertainty (5~mas for coronagraphic observations with SPHERE), plate scale error and true North error.
\end{table}

\begin{figure}
    \includegraphics[width=\columnwidth]{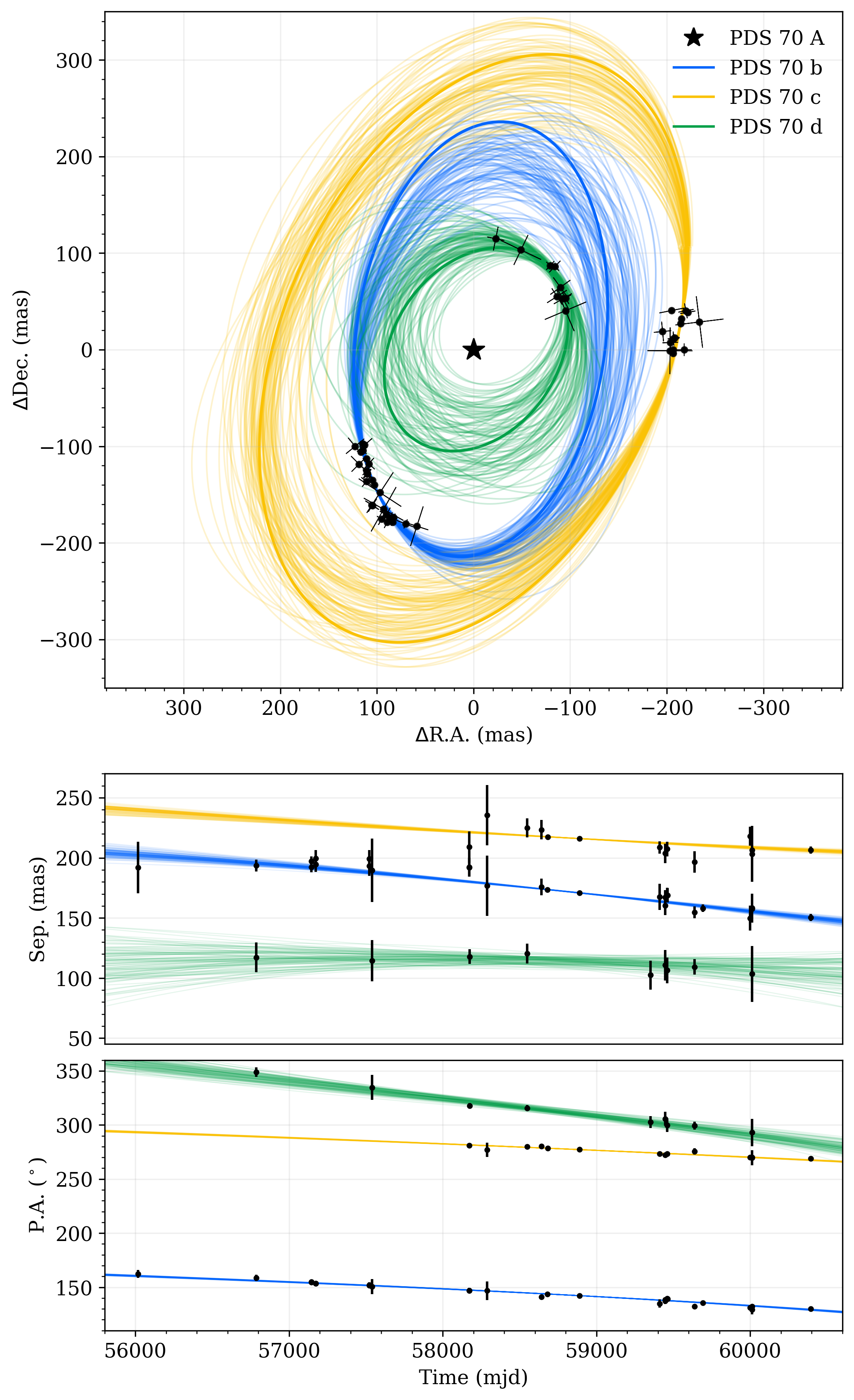}
    \caption{The maximum likelihood sample (solid) and 
    100 randomly drawn samples (transaprent) from the \textsc{Octofitter} orbital solutions for three protoplanets PDS~70~b (blue), PDS~70~c (orange) and candidate~d (green), without constraining to any resonances or enforcing coplanarity. \textit{Top:} Orbits as viewed on-sky, with our measured astrometry and associated error. 
    \textit{Bottom}: Separation and PA against time.
    }
    \label{fig:d_PA}
\end{figure}

\subsection{Orbital Predictions}\label{sec:astrometry}

\begin{figure*}
    \includegraphics[width=\textwidth]{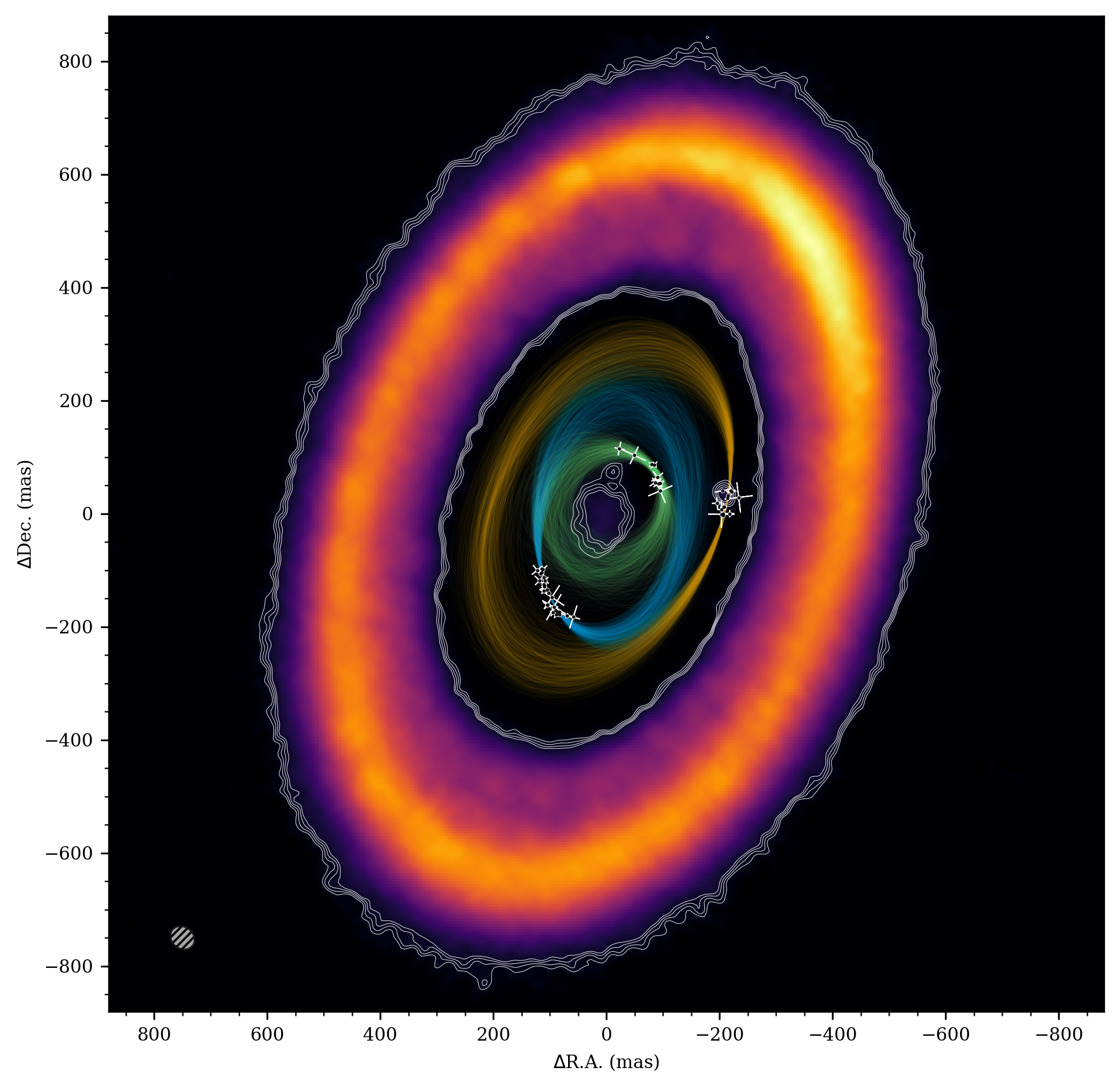}
    \caption{Overview of the PDS~70 system, including 1000 random samples from the orbital posteriors of planets b (blue curves), c (yellow), and candidate d (green). The priors and median posterior for each orbital element retrieved with \textsc{Octofitter} are summarized in Table \ref{tab:priors}. A shift in declination of $-0\farcs02$ was applied to the continuum emission (colourscale) to enforce the c\textsubscript{smm} blob to overlap with near-IR astrometry of the same epoch. The grey beam of 44 $\times$ 37~mas with $\mathrm{PA} = 51\fdg5$ in the bottom-left is for the continuum emission originally presented in \citet{Benisty:2021uj}.}
    \label{fig:three_body_problem}
\end{figure*}

We used \textsc{Octofitter}\footnote{\url{https://sefffal.github.io/Octofitter.jl}} v6.0.0 \citep{Thompson:2023a} to infer the posterior distribution of the orbital parameters for three simulated planets. We sample the posterior using stabilized variational non-
reversible parallel tempering \citep{10.1111/rssb.12464,Surjanovic:2022}, implemented in \textsc{Pigeons.jl} \citep{Surjanovic:2023}. Our choice of package was primarily motivated by computation time required to reach convergence and the ability to fit multiple planet orbits simultaneously, as \textsc{Octofitter} is the fastest orbital fitting code publicly available. We used eccentricity (\textit{e}), inclination (\textit{i}), semi-major axis (\textit{a}), argument of periastron (\textit{$\omega$}), longitude of the ascending node (\textit{$\Omega$}) 
and epoch of periastron, which is derived from the position angle at a reference epoch of MJD = 58889,
to define the orbit. For the b \& c planets, we used literature astrometry summarised in \citet{Wang:2021a}, with additional JWST/NIRISS, JWST/NIRCam and MagAO-X measurements \citep[][respectively]{Blakely:2025a, Christiaens:2024, Close:2025}.


We fit for all four bodies (A, b, c \& d) simultaneously and include planet--star and planet--planet interactions. To explore the parameter space for d, we set unconstrained distributions on all priors except for separation and mass, which were set to 1 -- 20~au and 0 -- 10$M_{\rm Jup}$, respectively. For the other planets, we implemented a prior on the mutual inclinations between b and the disc, c and the disc and b \& c, to be within $\pm$10$^{\circ}$. For the disc, we set \textit{i} = 128.3$^{\circ}$ and $\Omega = 156.7^{\circ}$ \citep{2019A&A...625A.118K}. A Gaussian prior centred at 0.88$M_{\odot}$ with a standard deviation of 0.09$M_{\odot}$ was used for the stellar mass, using the measurement derived by \cite{Wang:2021a}, including the same conservative 10$\%$ uncertainty as was used in their orbital analysis. 
Finally, we included Gaussian priors on the parallax and proper motion of the star from \textit{Gaia} DR3 \citep{Gaia-Collaboration:2023}. A summary of all priors is provided in Table \ref{tab:priors}. We also imposed that the orbits of b, c \& d are non-crossing, and the semi-major axis of c could not exceed 40~au based upon modelling of the outer dust ring in \citet{Zhou:2025}.
Figure~\ref{fig:d_PA} provides the orbital fit results for the three protoplanets obtained for the maximum likelihood sample, along with 100 randomly drawn posterior samples. Figure~\ref{fig:three_body_problem} shows the orbits retrieved for 1000 randomly drawn samples, and the corner plot is provided in Fig.~\ref{fig:corner_plot}. 

For the known protoplanets b \& c, we find similar, but notably different parameters to those reported by \citet{Wang:2021a}. We find a smaller stellar mass, along with a larger semi-major axis and lower eccentricity for planet b. To test if this can be explained by the presence of planet d, we compared our astrometry model with that of \citet{Wang:2021a} by testing a two planet model. We found near-identical numerical results between \textsc{orbitize!} and \textsc{Octofitter} for individual model evaluations, but were not able to replicate the same overall posterior distribution of the two-planet solution with either \textsc{Octofitter} with \textsc{Pigeons} or \textsc{orbitize!} with \textsc{ptemcee}. A future work could further investigate the source of this disagreement.

We were able to constrain the semi-major axis of PDS~70~d to 12.9$^{+2.0}_{-2.1}$~au when including all four bodies in the fit without enforcing any orbital resonances. Inclinations approximately equal to that of the outer dust continuum ring ($\sim$130\degr~in \citealt{Casassus:2022}) feature the strongest likelihood within our prior range for b \% c. The inclination and eccentricity of d, however, is not as constrained. Without the precision of VLTI/GRAVITY astrometry these orbital parameters will remain difficult to narrow down.

\begin{table}\centering
\caption{\textsc{Octofitter} orbital parameters and priors.}
\label{tab:priors}
\begin{tabular}{lccc}\hline\hline
Element & Prior & Prior Range & Median Posterior\\ \hline
$a_b$ (au) & Log Uniform & 1 -- 30 & $23.7^{+2.1}_{-2.0}$\\
$e_b$ & Uniform & 0 -- 0.99 & 0.12$^{+0.07}_{-0.07}$ \\
$i_b$ ($^{\circ}$) & Sine & $-180$ -- $+180$ & $128.1^{+3.7}_{-2.8}$\\
$\Omega_b$ ($^{\circ}$) & Uniform & 0 -- 360 & 171$^{+6}_{-7}$\\
$\omega_b$ ($^{\circ}$) & Uniform & 0 -- 360 & $110^{+47}_{-46}$\\
$\theta_{\textrm{b}}$ & Uniform & 0 -- 360 & 142.17$^{+0.03}_{-0.03}$\\
$M_{\textrm{b}}$ ($M_{\rm Jup}$) & Uniform & 1 -- 10 & 5.6$^{+3.0}_{-3.0}$ \\ 

\hline

$a_c$ (au) & Log Uniform & 1 -- 45 & 33.6$^{+2.4}_{-2.6}$ \\
$e_c$ & Uniform & 0 -- 0.99 & 0.05$^{+0.05}_{-0.03}$ \\
$i_c$ ($^{\circ}$) & Sine & $-180$ -- $+180$ &131.7$^{+3.2}_{-3.1}$\\
$\Omega_c$ ($^{\circ}$) & Uniform & 0 -- 360 & 155$^{+5}_{-6}$\\
$\omega_c$ ($^{\circ}$) & Uniform & 0 -- 360 & 185$^{+136}_{-156}$ \\
$\theta_{\textrm{c}}$ & Uniform & 0 -- 360 & 277.3$^{+0.1}_{-0.1}$ \\
$M_{\textrm{c}}$ ($M_{\rm Jup}$) & Uniform & 1 -- 10 & 5.4$^{+3.1}_{-3.1}$ \\

\hline
$a_d$ (au) & Log Uniform & 1 -- 20 & 12.9$^{+2.0}_{-2.1}$\\
$e_d$ & Uniform & 0 -- 0.99 & 0.16$^{+0.16}_{-0.12}$ \\
$i_d$ ($^{\circ}$) & Sine & $-180$ -- $+180$ & 146$^{+12}_{-9}$\\
$\Omega_d$ ($^{\circ}$) & Uniform & 0 -- 360 & 177$^{+145}_{-103}$\\
$\omega_d$ ($^{\circ}$) & Uniform & 0 -- 360 & 181 $^{+110}_{-122}$\\
$\theta_{\textrm{d}}$ & Uniform & 0 -- 360 & 310.1$^{+1.5}_{-1.6}$ \\
$M_{\textrm{d}}$ ($M_{\rm Jup}$) & Uniform & 0 -- 10 & 5.2$^{+3.3}_{-3.5}$ \\
\hline

$M_{\textrm{$*$}}$ ($M_\odot$) & Normal & $0.88 \pm 0.09$ & 0.92$^{+0.07}_{-0.07}$ \\ 
Parallax (mas) & Normal & $8.897\pm0.019$ & 8.90$^{+0.02}_{-0.02}$\\
\hline

\end{tabular}
Notes: A Gaussian co-planarity prior of $\pm10^{\circ}$ was placed on the mutual inclination of b with the disc, c with the disc and b with c. We imposed non-crossing orbits for all planets, and a constraint on c to not exceed 40~au. $\theta$ is the position angle at a reference epoch of MJD $=$ 58889.
\end{table}

\subsection{Spectrum}\label{sec:spectrum}
We opted to use the star-hopping strategy to retrieve the spectrum for the \textit{YJH} and \textit{Ks}--bands as this approach results in the fewest geometric biases at short separation from the star \citep{Wahhaj:2021, Xie:2022}. For simplicity we consider only the 2018 Feb 25 epoch in \emph{K$_1$K$_2$} as it was the longest sequence of its kind and was taken in the most favourable conditions. 

For IFS, we used the following technique to retrieve the spectrum: \textit{i}) measure the SNR of candidate~d in individual spectral channels (Fig.~\ref{fig:all_channels}) to determine the ten most favourable channels for astrometry, \textit{ii}) repeat the MCMC-NEGFC technique as in Section~\ref{sec:moving_source} after collapsing the ten best channels to infer the separation and PA of d, 
\textit{iii}) run a Nelder-Mead simplex-NEGFC over all 39 spectral channels individually by fitting the curvature of the pixel intensities in a 5 $\times$ 5 pixel subframe (as for NaCo, Section~\ref{sec:moving_source}) to retrieve the contrast of d, with the separation and PA fixed to the value found in step \textit{ii}, and \textit{iv}) inject fake companions at the same separation and flux as estimated for d, but at all azimuths in increments of 1\degr~excluding 10\degr~on either side of b and d, and estimate their retrieved flux after PCA-RDI using Nelder-Mead simplex-NEGFC with the \texttt{speckle\_noise\_uncertainty} function of \textsc{vip}. The PCA-RDI images for each IFS channel after subtracting our best flux estimate for d is shown in Fig. \ref{fig:sub_all_channels}. For IRDIS \emph{K$_1$} , \emph{K$_2$} and \textit{Ks} bands we directly took the resulting flux from the MCMC-NEGFC technique in Section \ref{sec:moving_source}, and performed the speckle noise uncertainty procedure to retrieve errors on the flux.

For NaCo we used the flux and associated error recovered from the Nelder-Mead simplex-NEGFC and speckle noise uncertainty, as described in Section \ref{sec:astrometry}. We employed the same technique for SINFONI, however identified the best fit negative flux by minimizing the standard deviation of residual intensities in a 0.7$\times$FWHM aperture around d rather than fitting the curvature of the pixel intensities.

To convert from detector units to physical units, we took the contrast ratio of d and that of PDS~70~A measured in the non-coronographic PSF frames, and multiplied it by the flux-calibrated spectrum of PDS~70~A acquired with SpeX at the NASA Infrared Telescope Facility (IRTF). This choice of spectrum was motived by the need to include unresolved hot inner disc signal. 
As the SpeX data do not extend to 3.78~\mum, we interpolated NASA/WISE W1 (3.35~\mum) and W2 (4.6~\mum) photometry in log space. For NIRCam we simply used the flux-calibrated F187N measurement of the star published in \citet{Christiaens:2024}. 

There are two caveats with the above technique. First, d is at a radial separation that is partially attenuated by the SPHERE coronagraph by $\sim$30\% and this attenuation varies with coronagraph azimuth (and consequently, by epoch). We used the transmission profile for the \texttt{N\_ALC\_YJH\_S} with the YJ prism as given in the SPHERE manual to scale the flux from d for each wavelength, which is not directly equivalent to the radial transmission expected for the YJH prism. Second, PDS~70~A was placed behind the coronagraphic mask for the FLUX cubes during the 2022 Feb 28 star-hopping epoch, meaning there was no direct observation of this star on this night. We instead used FLUX cubes taken in similar observing conditions and with the YJH prism from the 2021 May 17 observation for IFS and the 2021 Sep 03 observation for IRDIS, ensuring to correctly scale the DIT. Any variability of the star at these wavelengths between these observations would be folded into the spectrum.

\begin{figure*}
    \includegraphics[width=\textwidth]{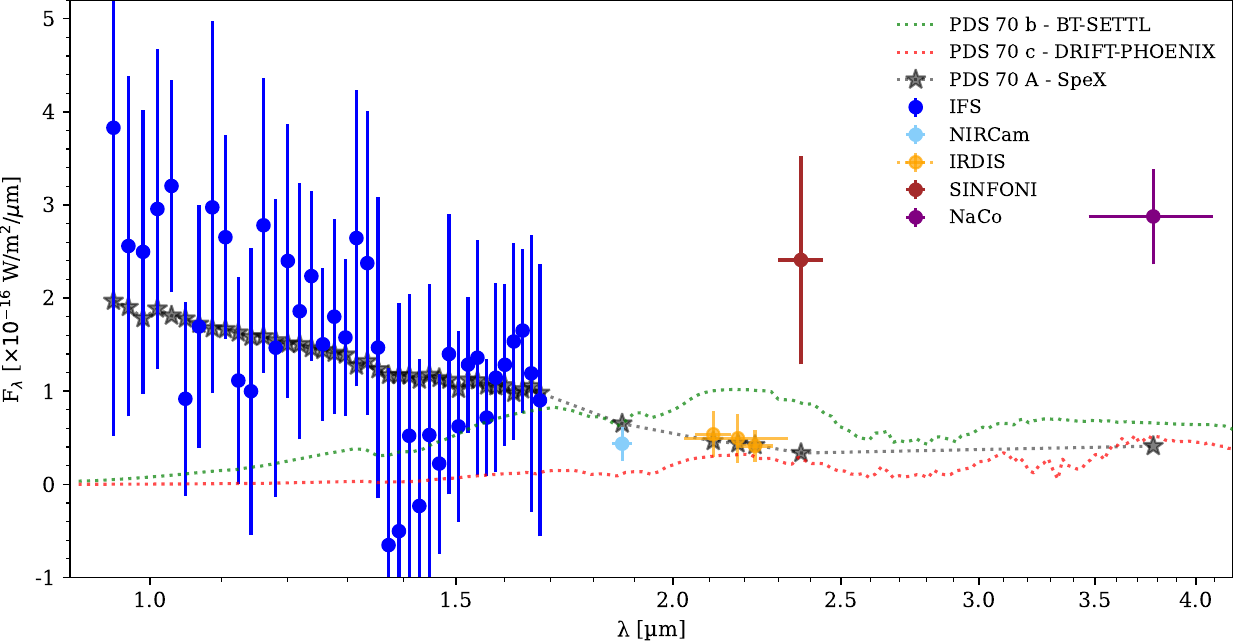}
    \caption{Extracted spectrum of the PDS~70~d candidate for all instruments presented in this work. For the 0.95--1.65~\mum (blue) and 2.18~\mum (orange) measurements we used the 2022 Feb 28 star-hopping epoch, while the 2.11~\mum and 2.23~\mum (orange) measurements are from 2018 Feb 25. The resampled stellar spectrum (gray) we used to convert to flux units is from SpeX at the IRFT and was scaled by 3$\times$10$^{-4}$ to match the intensity of candidate~d at recent epochs. Horizontal error bars correspond to the effective filter width for each instrument. The spectral slope of the candidate is overall consistent with that of the star, except for the SINFONI and NaCo points for which possible explanations are discussed in text. 
    }
    \label{fig:ifs_spectrum}
\end{figure*}

Figure~\ref{fig:ifs_spectrum} shows the spectrum and associated uncertainty. The shape is consistent with the SED of the star and unresolved inner disc spectrum at short wavelengths for the RDI strategy, however the uncertainty due to speckle noise is significant at the separation of the source. The NIRCam and IRDIS measurements follow the same trend as the IFS, whereas the longer wavelength SINFONI and NaCo measurements are significantly brighter. 

Several mechanisms can produce the high flux we measured at 2.37~\mum and 3.78~\mum. PDS~70~A is known to be highly variable \citep{Perotti:2023, Gaidos:2024, Jang:2024} likely due to occulting material close to the star. This is also indicated by the WISE W1 (3.35~\mum) time-series observations, which is the closest WISE band to our \emph{L'} measurement. If the star and unresolved inner disc were dimmer during the SINFONI and NaCo epochs, we would overestimate the 2.37~\mum and 3.78~\mum flux. For this reason, we corrected all of our flux measurements for the stellar variability measured in WISE W1 by assuming a consistent correlation for all wavelengths. The largest change applied to the SINFONI epoch, where PDS 70~A dimmed by $\sim$40\% at 3.35~\mum (see Fig.~7 of \citealt{Gaidos:2024}). This may nonetheless not be a sufficient correction given the sparse sampling of stellar variability and the assumed linear relation between measurements, which may underpredict the flux drop at the SINFONI epoch if already undergoing the dipper event. Another source of bias is the inclusion of an unknown amount of contamination from the inner disc in the PSF. This is most relevant for NaCo where the FWHM during the sequence was 108~mas, although may also be relevant to the SINFONI data where the achieved Strehl ratio is likely lower than in the SPHERE data. 



\section{Discussion}
\label{sec:discussion}
\subsection{Evidence for a third protoplanet}
The source is detected across nine epochs over nine years. This lends support to it being true signal rather than a post-processing artefact, and can rule out the hypothesis of the signal tracing a static inner disc or residual speckle. A speckle would be transient across observations and not be consistent through the instruments included in this work. Candidate~d is also detected in individual spectral channels (Appendix~\ref{sec:allchans}). Additionally, the source is not in the predicted location of the Lagrange points L$_4$ or L$_5$ for either b or c \citep{Balsalobre-Ruza:2023}. It is therefore not expected to be tracing a dust trap related to these protoplanets. Additionally, \textsc{mayonnaise} identifies it as a point source rather than a component of the disc (Appendix~\ref{sec:andromeda+mayo_data}) although there is another feature at the same separation pointed out as PLF~2 in \citet{Pairet:2021}. Note that we also detected d in the IRDIS 2018 Feb 25 data with the \textsc{andromeda} \citep{Cantalloube:2015wa} post-processing algorithm, which filters out extended structures to find point-like sources (Appendix~\ref{sec:andromeda+mayo_data}). While not terribly revealing, a third body easily fits within the observed astrometric Reduced Unit Weight Error (RUWE) reported from \textit{Gaia} (Appendix~\ref{sec:ruwe}).


Orbital fitting in Section \ref{sec:astrometry} shows that d can be described as a bound object with Keplerian frequency approximately in the plane of the disc - as are b \& c. Interestingly, the favoured solutions allude to a Laplace resonance with the two known protoplanets. Considering the uncertainties on the period of candidate d and on the stellar mass, the exact resonance is not clear. The median posterior from \textsc{Octofitter} gives a period ratio of d \& b as {2.46$^{+0.82}_{-0.53}$, also including the possibility of the 3:1 or 5:2 resonance, and of c \& d it is 4.16 $^{+1.32}_{-0.93}$. Using the dynamically stable result for b in \citet{Wang:2021a} gives a period ratio of d \& b of 2.05$^{+0.74}_{-0.48}$. Given the mass of the other bodies we expect such a MMR in order for a chain of three giant planets to be stable over long time periods (Myr). As the age of PDS~70~A is 5.4$\pm$1 Myr \citep{Muller:2018wg} any protoplanets would need to be in strong, preferably first order, resonance to avoid ejection. Such a scenario would be similar to the directly imaged giant planets HR~8799 b, c \& d \citep{Gozdziewski:2020}, the Gliese-876 b, c and d giant planets \citep{Nelson:2016, Cimerman:2018}, and the Galilean moons, which all share a Laplace resonance. 

An explanation for the SINFONI flux could be the first CO overtone \textit{v} = 0 to \textit{v} = 2 which occurs at 2.38~\mum. Such excitation would require warm gas in the vicinity of a planet, heated by stellar radiation \citep{Oberg:2020}, by radiation from the planet, and by accretion. Similar predictions have been made for $^{12}$CO rotational–vibrational lines in \citet{Oberg:2023}. Alternatively, it may probe thermal excess towards the end of the \textit{K}-band from a circumplanetary disc (CPD). Interestingly, the flux estimated in the NaCo data lends support to the CPD hypothesis. This would also be consistent with b \& c where circumplanetary contribution is required to match the SED of the planets at these wavelengths \citep{Christiaens:2019vd, Blakely:2025a, Choksi:2025}.

The extended nature of d in the RDI observations is suggestive of a spiral wake or circumplanetary envelope, if d is a protoplanet. A protoplanet would dynamically interact with the inner disc and release spiral density waves \citep{Ogilvie:2002tq}. In this scenario, the signal seen at short wavelengths could therefore be starlight scattering off small grains in the outer wake of the protoplanet. 


\subsection{Evidence against a third protoplanet}
If indeed d is a protoplanet at a deprojected separation of 12.2~au, it is in the same annular gap as the other protoplanets. However, the extent of the inner disc is not well constrained and needs to be considered.
Recent ALMA 855~\mum observations in \citet{Benisty:2021uj} show a mostly resolved inner disc, and at higher resolutions its distribution is irregular. The ALMA data shows the presence of mm-emitting grains close to the separation of candidate d. However the orbital distance of the mm-emitting grains shrinks with a decreasing \texttt{robust} parameter (\texttt{robust} = 0.5 in our Fig. \ref{fig:three_body_problem}), and an inner disc is not detected in the 3~mm continuum \citep{Doi:2024}. \citet{Benisty:2021uj} modelled the 855~\mum signal using a Gaussian with a width of 59~mas. By taking 2$\sigma$ (twice the width of the Gaussian) we get an inner disc size of about 120~mas, which is 60~mas each side of the star (7~au projected). Candidate~d is exterior ($\sim$110 mas) to such an inner disc assuming a uniform distribution of grains. 

By paying special attention to uncertainties and the JvM correction \citep{Jorsater:1995} of the same 855~\mum data, \citet{Casassus:2022} further unveiled the finer structure of the inner disc. It is highly irregular and the morphology varies significantly over three epochs (2016, 2017 and 2019). Most notably, there is a detached clump of emission towards the North at 0\farcs08, or 4.5~au projected, separation from the ring centre. Such structures seem to change between epochs, indicating they are either variable or artefacts that require further study. We stress that such emission is interior to d observed in the near-IR, although the presence of variable, ambient material in proximity complicates any claim that d could be a protoplanet and detached from the inner disc. If it is physical, what drives this irregular structure is unclear, but could be related to the gravitational perturbation from a nearby protoplanet.


Many of the orbital solutions in Fig. \ref{fig:three_body_problem} pass close to the 855~\mum inner disc continuum with a beam of 44 $\times$ 37~mas. A protoplanet of similar mass to PDS~70~b \& c should have carved out a dust gap in its orbital path. Whether the irregular and clumpy distribution of the inner disc can be attributed to dynamical interaction with a protoplanet at $\sim13$~au, or simply to b alone, is difficult to disentangle. New hydrodynamical simulations with the addition of a third protoplanet at the separation of d could indicate whether the inner disc is stable over several Myr with such a configuration of protoplanets. 


\subsection{Implications for future observations}
Based on our fits for a Keplerian protoplanet we expect planet~d to orbit clockwise and reach the semi-minor axis of the disc within a few years. For this reason, its polarised intensity will decrease due to the scattering angle dependence (forward vs. backward scattering), whereas the total intensity will increase. While this is favourable for future studies, the predicted orbits indicate that d will pass further behind a coronagraphic mask due to the viewing angle on-sky. This caveat may motivate non-coronagraphic follow-up campaigns, particularly with star-hopping, to reach deeper contrast at short separation. Such an observation would need to be taken in more stable observing conditions than previously due to the presence of bright speckles at close separations. 

Follow-up observations with VLTI/GRAVITY are an obvious next step to significantly constrain the orbit of d, as has already been demonstrated with two epochs on both b \& c \citep{Wang:2021a}. The detection of d in \textit{K}-bands with SPHERE/IRDIS suggests that it will be detectable in the operating wavelengths of GRAVITY (2.0--2.4~\mum) and such an observation would provide a higher resolution spectrum. The $\sim$50 mas field of view of GRAVITY is wide enough to capture all orbital solutions found with existing HCI in this work for at least the next few years. 
However the presence of bright extended signals, such as those present in the star-hopping data in the vicinity of d, could complicate an observation with GRAVITY. 

Thermal emission, or lack thereof, in ELT/METIS observations may be required to fully distinguish between the hypothesis of d being a disc signal or a third protoplanet, as thermal contribution is expected to come either directly from the surface of the planet or from the CPD \citep{Szulagyi:2019vo, Choksi:2025}. The higher angular resolution of the ELT at these wavelengths would disentangle inner disc signal from any real planet signal and enable more robust spectral characterisation.

\section{Conclusions}
\label{sec:conclusion}
We present nine years of high-contrast imaging data of PDS~70 with 
VLT/SINFONI, VLT/NaCo, VLT/SPHERE, and JWST/NIRCam to obtain estimates on the orbit and spectrum of a third protoplanet candidate. Our analysis of ten data sets over eight epochs, plus the published JWST/NIRCam observations, suggest the following:
\begin{enumerate}
    \item The candidate protoplanet PDS 70~d is detected at 0.95--3.8~\mum, irrespective of observing strategy (ADI, SDI or RDI) or post-processing technique, in clockwise motion consistent with \citet{Mesa:2019a} and \citet{Christiaens:2024}.
    \item Orbital solutions suggest a near 4:2:1 Laplace resonance with the existing protoplanets in the plane of the disc, although the eccentricity is uncertain. The semi-major axis of d is 12.9$^{+2.0}_{-2.1}$~au when including all four bodies in the fit (A, b, c, d).
    \item The \textit{YJH} spectrum of d is consistent with the SED of the star and unresolved inner disc, suggestive of starlight scattering off small dust grains and is also bluer than planet b.
    \item A thermal excess is detected at 2.37~\mum and 3.78~\mum. Its origin is complicated by stellar variability and contamination from the inner disc, but such an excess can come from a CPD. 
    \item The source may not be `point-like' as previously reported, and instead appears extended in RDI data due to the absence of ADI self-subtraction. To explain such morphology, the feature could be tracing an outer spiral wake from a third protoplanet with a dusty envelope, and could explain the lack of accretion signatures.
    \item The extent of the inner disc around PDS~70~A is not well constrained and, in some cases, is close to the predicted orbits of PDS~70~d. 
\end{enumerate}

\section*{Acknowledgements}
We thank Myriam Benisty for useful discussions and for kindly supplying the ALMA data. We also thank Yifan Zhou for discussions regarding the HST observations. IH acknowledges a Research Training Program scholarship from the Australian government. VC and SJ thank the Belgian F.R.S.-FNRS. VC thanks the Belgian Federal Science Policy Office (BELSPO) for the provision of financial support in the framework of the PRODEX Programme of the European Space Agency (ESA) under contract number 4000142531. GDM acknowledges the support from the European Research Council (ERC) under the Horizon 2020 Framework Program via the ERC Advanced Grant ``ORIGINS'' (PI: Henning), Nr.~832428, and via the research and innovation programme ``PROTOPLANETS'', grant agreement Nr.~101002188 (PI: Benisty). This work is based in part on observations made with the NASA/ESA/CSA James Webb Space Telescope. The data were obtained from the Mikulski Archive for Space Telescopes at the Space Telescope Science Institute, which is operated by the Association of Universities for Research in Astronomy, Inc., under NASA contract NAS 5-03127 for JWST. These observations are associated with program 1282. We made use of the Multi-modal Australian ScienceS Imaging and Visualisation Environment (MASSIVE) and data from the European Space Agency (ESA) mission \textit{Gaia}. This work was performed on the OzSTAR national facility at Swinburne University of Technology. The OzSTAR program receives funding in part from the Astronomy National Collaborative Research Infrastructure Strategy (NCRIS) allocation provided by the Australian Government, and from the Victorian Higher Education State Investment Fund (VHESIF) provided by the Victorian Government. We are grateful for Australian Research Council Discovery Project funding via DP220103767 and DP240103290.\\
The scientific results presented in this article made use of Python and the \textsc{numpy}, \textsc{scipy}, \textsc{astropy} and \textsc{matplotlib} packages.

\section*{Data Availability}
Data are available as FITS files through the Max Planck Digital Library’s \textit{Keeper} research
repository via the following link: \href{https://keeper.mpdl.mpg.de/d/2633616c51b64433962d/}{https://keeper.mpdl.mpg.de/d/2633616c51b64433962d/}



\bibliographystyle{mnras}
\bibliography{references} 

\begin{thebibliography}{}
\makeatletter
\relax
\def\mn@urlcharsother{\let\do\@makeother \do\$\do\&\do\#\do\^\do\_\do\%\do\~}
\def\mn@doi{\begingroup\mn@urlcharsother \@ifnextchar [ {\mn@doi@} {\mn@doi@[]}}
\def\mn@doi@[#1]#2{\def\@tempa{#1}\ifx\@tempa\@empty \href {http://dx.doi.org/#2} {doi:#2}\else \href {http://dx.doi.org/#2} {#1}\fi \endgroup}
\def\mn@eprint#1#2{\mn@eprint@#1:#2::\@nil}
\def\mn@eprint@arXiv#1{\href {http://arxiv.org/abs/#1} {{\tt arXiv:#1}}}
\def\mn@eprint@dblp#1{\href {http://dblp.uni-trier.de/rec/bibtex/#1.xml} {dblp:#1}}
\def\mn@eprint@#1:#2:#3:#4\@nil{\def\@tempa {#1}\def\@tempb {#2}\def\@tempc {#3}\ifx \@tempc \@empty \let \@tempc \@tempb \let \@tempb \@tempa \fi \ifx \@tempb \@empty \def\@tempb {arXiv}\fi \@ifundefined {mn@eprint@\@tempb}{\@tempb:\@tempc}{\expandafter \expandafter \csname mn@eprint@\@tempb\endcsname \expandafter{\@tempc}}}

\bibitem[\protect\citeauthoryear{{Absil} et~al.,}{{Absil} et~al.}{2013}]{Absil:2013}
{Absil} O.,  et~al., 2013, \mn@doi [\aap] {10.1051/0004-6361/201322748}, \href {https://ui.adsabs.harvard.edu/abs/2013A&A...559L..12A} {559, L12}

\bibitem[\protect\citeauthoryear{{Amara} \& {Quanz}}{{Amara} \& {Quanz}}{2012}]{Amara:2012um}
{Amara} A.,  {Quanz} S.~P.,  2012, \mn@doi [\mnras] {10.1111/j.1365-2966.2012.21918.x}, \href {https://ui.adsabs.harvard.edu/abs/2012MNRAS.427..948A} {427, 948}

\bibitem[\protect\citeauthoryear{{Bae} et~al.,}{{Bae} et~al.}{2019}]{Bae:2019}
{Bae} J.,  et~al., 2019, \mn@doi [\apjl] {10.3847/2041-8213/ab46b0}, \href {https://ui.adsabs.harvard.edu/abs/2019ApJ...884L..41B} {884, L41}

\bibitem[\protect\citeauthoryear{{Balsalobre-Ruza} et~al.,}{{Balsalobre-Ruza} et~al.}{2023}]{Balsalobre-Ruza:2023}
{Balsalobre-Ruza} O.,  et~al., 2023, \mn@doi [\aap] {10.1051/0004-6361/202346493}, \href {https://ui.adsabs.harvard.edu/abs/2023A&A...675A.172B} {675, A172}

\bibitem[\protect\citeauthoryear{{Benisty} et~al.,}{{Benisty} et~al.}{2021}]{Benisty:2021uj}
{Benisty} M.,  et~al., 2021, \mn@doi [\apjl] {10.3847/2041-8213/ac0f83}, \href {https://ui.adsabs.harvard.edu/abs/2021ApJ...916L...2B} {916, L2}

\bibitem[\protect\citeauthoryear{{Beuzit} et~al.,}{{Beuzit} et~al.}{2008}]{Beuzit:2008}
{Beuzit} J.-L.,  et~al., 2008, in {McLean} I.~S.,  {Casali} M.~M.,  eds,  SPIE Conf. Ser. Vol. 7014, Ground-based \& Airborne Instr. for Astronomy II. p. 701418

\bibitem[\protect\citeauthoryear{{Blakely} et~al.,}{{Blakely} et~al.}{2025}]{Blakely:2025a}
{Blakely} D.,  et~al., 2025, \mn@doi [\aj] {10.3847/1538-3881/ad9b94}, \href {https://ui.adsabs.harvard.edu/abs/2025AJ....169..137B} {169, 137}

\bibitem[\protect\citeauthoryear{{Cantalloube} et~al.,}{{Cantalloube} et~al.}{2015}]{Cantalloube:2015wa}
{Cantalloube} F.,  et~al., 2015, \mn@doi [\aap] {10.1051/0004-6361/201425571}, \href {https://ui.adsabs.harvard.edu/abs/2015A&A...582A..89C} {582, A89}

\bibitem[\protect\citeauthoryear{{Casassus} \& {C{\'a}rcamo}}{{Casassus} \& {C{\'a}rcamo}}{2022}]{Casassus:2022}
{Casassus} S.,  {C{\'a}rcamo} M.,  2022, \mn@doi [\mnras] {10.1093/mnras/stac1285}, \href {https://ui.adsabs.harvard.edu/abs/2022MNRAS.513.5790C} {513, 5790}

\bibitem[\protect\citeauthoryear{{Choksi} \& {Chiang}}{{Choksi} \& {Chiang}}{2025}]{Choksi:2025}
{Choksi} N.,  {Chiang} E.,  2025, \mn@doi [\mnras] {10.1093/mnras/stae2530}, \href {https://ui.adsabs.harvard.edu/abs/2025MNRAS.537.2945C} {537, 2945}

\bibitem[\protect\citeauthoryear{{Christiaens} et~al.,}{{Christiaens} et~al.}{2019a}]{Christiaens:2019}
{Christiaens} V.,  et~al., 2019a, \mn@doi [\mnras] {10.1093/mnras/stz1232}, \href {https://ui.adsabs.harvard.edu/abs/2019MNRAS.486.5819C} {486, 5819}

\bibitem[\protect\citeauthoryear{{Christiaens} et~al.}{{Christiaens} et~al.}{2019b}]{Christiaens:2019vd}
{Christiaens} V.,  et~al., 2019b, \mn@doi [\apjl] {10.3847/2041-8213/ab212b}, \href {https://ui.adsabs.harvard.edu/abs/2019ApJ...877L..33C} {877, L33}

\bibitem[\protect\citeauthoryear{{Christiaens} et~al.,}{{Christiaens} et~al.}{2021}]{Christiaens:2021}
{Christiaens} V.,  et~al., 2021, \mn@doi [\mnras] {10.1093/mnras/stab480}, \href {https://ui.adsabs.harvard.edu/abs/2021MNRAS.502.6117C} {502, 6117}

\bibitem[\protect\citeauthoryear{{Christiaens}, {Hammond}, {Juillard}, {Kokoulina}  \& {Balsalobre-Ruza}}{{Christiaens} et~al.}{2023a}]{Christiaens:2023a}
{Christiaens} V.,  {Hammond} I.,  {Juillard} S.,  {Kokoulina} E.,   {Balsalobre-Ruza} O.,  2023a, {VCAL-SPHERE: Hybrid pipeline for reduction of VLT/SPHERE data}, Astrophysics Source Code Library, record ascl:2311.002 (\mn@eprint {ascl} {2311.002})

\bibitem[\protect\citeauthoryear{{Christiaens} et~al.,}{{Christiaens} et~al.}{2023b}]{Christiaens:2023}
{Christiaens} V.,  et~al., 2023b, \mn@doi [The Journal of Open Source Software] {10.21105/joss.04774}, \href {https://ui.adsabs.harvard.edu/abs/2023JOSS....8.4774C} {8, 4774}

\bibitem[\protect\citeauthoryear{{Christiaens} et~al.,}{{Christiaens} et~al.}{2024}]{Christiaens:2024}
{Christiaens} V.,  et~al., 2024, \mn@doi [\aap] {10.1051/0004-6361/202349089}, \href {https://ui.adsabs.harvard.edu/abs/2024A&A...685L...1C} {685, L1}

\bibitem[\protect\citeauthoryear{{Cimerman}, {Kley}  \& {Kuiper}}{{Cimerman} et~al.}{2018}]{Cimerman:2018}
{Cimerman} N.~P.,  {Kley} W.,   {Kuiper} R.,  2018, \mn@doi [\aap] {10.1051/0004-6361/201833591}, \href {https://ui.adsabs.harvard.edu/abs/2018A&A...618A.169C} {618, A169}

\bibitem[\protect\citeauthoryear{{Close} et~al.,}{{Close} et~al.}{2025}]{Close:2025}
{Close} L.~M.,  et~al., 2025, \mn@doi [\aj] {10.3847/1538-3881/ad8648}, \href {https://ui.adsabs.harvard.edu/abs/2025AJ....169...35C} {169, 35}

\bibitem[\protect\citeauthoryear{{Desidera} et~al.,}{{Desidera} et~al.}{2021}]{Desidera:2021}
{Desidera} S.,  et~al., 2021, \mn@doi [\aap] {10.1051/0004-6361/202038806}, \href {https://ui.adsabs.harvard.edu/abs/2021A&A...651A..70D} {651, A70}

\bibitem[\protect\citeauthoryear{{Doi}, {Kataoka}, {Liu}, {Yoshida}, {Benisty}, {Dong}, {Yamato}  \& {Hashimoto}}{{Doi} et~al.}{2024}]{Doi:2024}
{Doi} K.,  {Kataoka} A.,  {Liu} H.~B.,  {Yoshida} T.~C.,  {Benisty} M.,  {Dong} R.,  {Yamato} Y.,   {Hashimoto} J.,  2024, \mn@doi [\apjl] {10.3847/2041-8213/ad7f51}, \href {https://ui.adsabs.harvard.edu/abs/2024ApJ...974L..25D} {974, L25}

\bibitem[\protect\citeauthoryear{{Fabrycky} \& {Murray-Clay}}{{Fabrycky} \& {Murray-Clay}}{2010}]{Fabrycky:2010}
{Fabrycky} D.~C.,  {Murray-Clay} R.~A.,  2010, \mn@doi [\apj] {10.1088/0004-637X/710/2/1408}, \href {https://ui.adsabs.harvard.edu/abs/2010ApJ...710.1408F} {710, 1408}

\bibitem[\protect\citeauthoryear{{Gaia Collaboration} et~al.,}{{Gaia Collaboration} et~al.}{2023}]{Gaia-Collaboration:2023}
{Gaia Collaboration} et~al., 2023, \mn@doi [\aap] {10.1051/0004-6361/202243940}, \href {https://ui.adsabs.harvard.edu/abs/2023A&A...674A...1G} {674, A1}

\bibitem[\protect\citeauthoryear{{Gaidos}, {Thanathibodee}, {Hoffman}, {Ong}, {Hinkle}, {Shappee}  \& {Banzatti}}{{Gaidos} et~al.}{2024}]{Gaidos:2024}
{Gaidos} E.,  {Thanathibodee} T.,  {Hoffman} A.,  {Ong} J.,  {Hinkle} J.,  {Shappee} B.~J.,   {Banzatti} A.,  2024, \mn@doi [\apj] {10.3847/1538-4357/ad3447}, \href {https://ui.adsabs.harvard.edu/abs/2024ApJ...966..167G} {966, 167}

\bibitem[\protect\citeauthoryear{{Gomez Gonzalez} et~al.,}{{Gomez Gonzalez} et~al.}{2017}]{Gomez-Gonzalez:2017uw}
{Gomez Gonzalez} C.~A.,  et~al., 2017, \mn@doi [\aj] {10.3847/1538-3881/aa73d7}, \href {https://ui.adsabs.harvard.edu/abs/2017AJ....154....7G} {154, 7}

\bibitem[\protect\citeauthoryear{{Go{\'z}dziewski} \& {Migaszewski}}{{Go{\'z}dziewski} \& {Migaszewski}}{2020}]{Gozdziewski:2020}
{Go{\'z}dziewski} K.,  {Migaszewski} C.,  2020, \mn@doi [\apjl] {10.3847/2041-8213/abb881}, \href {https://ui.adsabs.harvard.edu/abs/2020ApJ...902L..40G} {902, L40}

\bibitem[\protect\citeauthoryear{{Haffert}, {Bohn}, {de Boer}, {Snellen}, {Brinchmann}, {Girard}, {Keller}  \& {Bacon}}{{Haffert} et~al.}{2019}]{Haffert:2019td}
{Haffert} S.~Y.,  {Bohn} A.~J.,  {de Boer} J.,  {Snellen} I.~A.~G.,  {Brinchmann} J.,  {Girard} J.~H.,  {Keller} C.~U.,   {Bacon} R.,  2019, \mn@doi [Nature Astronomy] {10.1038/s41550-019-0780-5}, \href {https://ui.adsabs.harvard.edu/abs/2019NatAs...3..749H} {3, 749}

\bibitem[\protect\citeauthoryear{{Hammond}, {Christiaens}, {Price}, {Toci}, {Pinte}, {Juillard}  \& {Garg}}{{Hammond} et~al.}{2023}]{Hammond:2023}
{Hammond} I.,  {Christiaens} V.,  {Price} D.~J.,  {Toci} C.,  {Pinte} C.,  {Juillard} S.,   {Garg} H.,  2023, \mn@doi [\mnras] {10.1093/mnrasl/slad027}, \href {https://ui.adsabs.harvard.edu/abs/2023MNRAS.522L..51H} {522, L51}

\bibitem[\protect\citeauthoryear{{Hashimoto} et~al.,}{{Hashimoto} et~al.}{2012}]{Hashimoto:2012wi}
{Hashimoto} J.,  et~al., 2012, \mn@doi [\apjl] {10.1088/2041-8205/758/1/L19}, \href {https://ui.adsabs.harvard.edu/abs/2012ApJ...758L..19H} {758, L19}

\bibitem[\protect\citeauthoryear{{Hunziker}, {Quanz}, {Amara}  \& {Meyer}}{{Hunziker} et~al.}{2018}]{Hunziker:2018ux}
{Hunziker} S.,  {Quanz} S.~P.,  {Amara} A.,   {Meyer} M.~R.,  2018, \mn@doi [\aap] {10.1051/0004-6361/201731428}, \href {https://ui.adsabs.harvard.edu/abs/2018A&A...611A..23H} {611, A23}

\bibitem[\protect\citeauthoryear{{Jang} et~al.,}{{Jang} et~al.}{2024}]{Jang:2024}
{Jang} H.,  et~al., 2024, \mn@doi [\aap] {10.1051/0004-6361/202451589}, \href {https://ui.adsabs.harvard.edu/abs/2024A&A...691A.148J} {691, A148}

\bibitem[\protect\citeauthoryear{{Jorsater} \& {van Moorsel}}{{Jorsater} \& {van Moorsel}}{1995}]{Jorsater:1995}
{Jorsater} S.,  {van Moorsel} G.~A.,  1995, \mn@doi [\aj] {10.1086/117668}, \href {https://ui.adsabs.harvard.edu/abs/1995AJ....110.2037J} {110, 2037}

\bibitem[\protect\citeauthoryear{{Juillard}, {Christiaens}  \& {Absil}}{{Juillard} et~al.}{2022}]{Juillard:2022}
{Juillard} S.,  {Christiaens} V.,   {Absil} O.,  2022, \mn@doi [\aap] {10.1051/0004-6361/202244402}, \href {https://ui.adsabs.harvard.edu/abs/2022A&A...668A.125J} {668, A125}

\bibitem[\protect\citeauthoryear{{Juillard}, {Christiaens}  \& {Absil}}{{Juillard} et~al.}{2023}]{Juillard:2023}
{Juillard} S.,  {Christiaens} V.,   {Absil} O.,  2023, \mn@doi [\aap] {10.1051/0004-6361/202347259}, \href {https://ui.adsabs.harvard.edu/abs/2023A&A...679A..52J} {679, A52}

\bibitem[\protect\citeauthoryear{{Juillard}, {Christiaens}, {Absil}, {Stasevic}  \& {Milli}}{{Juillard} et~al.}{2024}]{Juillard:2024}
{Juillard} S.,  {Christiaens} V.,  {Absil} O.,  {Stasevic} S.,   {Milli} J.,  2024, \mn@doi [\aap] {10.1051/0004-6361/202449747}, \href {https://ui.adsabs.harvard.edu/abs/2024A&A...688A.185J} {688, A185}

\bibitem[\protect\citeauthoryear{{Keppler} et~al.,}{{Keppler} et~al.}{2018}]{Keppler:2018vt}
{Keppler} M.,  et~al., 2018, \mn@doi [\aap] {10.1051/0004-6361/201832957}, \href {https://ui.adsabs.harvard.edu/abs/2018A&A...617A..44K} {617, A44}

\bibitem[\protect\citeauthoryear{{Keppler} et~al.,}{{Keppler} et~al.}{2019}]{2019A&A...625A.118K}
{Keppler} M.,  et~al., 2019, \mn@doi [\aap] {10.1051/0004-6361/201935034}, \href {https://ui.adsabs.harvard.edu/abs/2019A&A...625A.118K} {625, A118}

\bibitem[\protect\citeauthoryear{{Kiefer}, {Bohn}, {Quanz}, {Kenworthy}  \& {Stolker}}{{Kiefer} et~al.}{2021}]{Kiefer:2021}
{Kiefer} S.,  {Bohn} A.~J.,  {Quanz} S.~P.,  {Kenworthy} M.,   {Stolker} T.,  2021, \mn@doi [\aap] {10.1051/0004-6361/202140285}, \href {https://ui.adsabs.harvard.edu/abs/2021A&A...652A..33K} {652, A33}

\bibitem[\protect\citeauthoryear{{Krijt}, {Kama}, {McClure}, {Teske}, {Bergin}, {Shorttle}, {Walsh}  \& {Raymond}}{{Krijt} et~al.}{2023}]{Krijt:2023}
{Krijt} S.,  {Kama} M.,  {McClure} M.,  {Teske} J.,  {Bergin} E.~A.,  {Shorttle} O.,  {Walsh} K.~J.,   {Raymond} S.~N.,  2023, in {Inutsuka} S.,  {Aikawa} Y.,  {Muto} T.,  {Tomida} K.,   {Tamura} M.,  eds,  Astronomical Society of the Pacific Conference Series Vol. 534, Protostars and Planets VII. p.~1031

\bibitem[\protect\citeauthoryear{{Lagrange} et~al.,}{{Lagrange} et~al.}{2010}]{Lagrange:2010}
{Lagrange} A.~M.,  et~al., 2010, \mn@doi [Science] {10.1126/science.1187187}, \href {https://ui.adsabs.harvard.edu/abs/2010Sci...329...57L} {329, 57}

\bibitem[\protect\citeauthoryear{Laplace}{Laplace}{1799}]{Laplace1799}
Laplace P.-S.,  1799, Traité de mécanique céleste.
~ Vol. 1-5, Crapart, Caille and Courcier, Paris

\bibitem[\protect\citeauthoryear{{Launhardt} et~al.,}{{Launhardt} et~al.}{2020}]{Launhardt:2020wu}
{Launhardt} R.,  et~al., 2020, \mn@doi [\aap] {10.1051/0004-6361/201937000}, \href {https://ui.adsabs.harvard.edu/abs/2020A&A...635A.162L} {635, A162}

\bibitem[\protect\citeauthoryear{Lindegren}{Lindegren}{2018}]{2018RenormalisingTA}
Lindegren L.,  2018, {Re-normalising the astrometric chi-square in Gaia DR2}, \url {https://api.semanticscholar.org/CorpusID:195836829}

\bibitem[\protect\citeauthoryear{{Long} et~al.,}{{Long} et~al.}{2018}]{Long:2018}
{Long} Z.~C.,  et~al., 2018, \mn@doi [\apj] {10.3847/1538-4357/aaba7c}, \href {https://ui.adsabs.harvard.edu/abs/2018ApJ...858..112L} {858, 112}

\bibitem[\protect\citeauthoryear{{Maire} et~al.,}{{Maire} et~al.}{2016}]{Maire:2016vq}
{Maire} A.-L.,  et~al., 2016, in {Evans} C.~J.,  et~al., eds,  SPIE Conf. Ser. Vol. 9908, Ground-based \& Airborne Instrumentation for Astronomy VI. p. 990834

\bibitem[\protect\citeauthoryear{{Marois} et~al.}{{Marois} et~al.}{2006a}]{Marois:2006vd}
{Marois} C.,  et~al., 2006a, \mn@doi [\apj] {10.1086/500401}, \href {https://ui.adsabs.harvard.edu/abs/2006ApJ...641..556M} {641, 556}

\bibitem[\protect\citeauthoryear{{Marois}, {Lafreni{\`e}re}, {Macintosh}  \& {Doyon}}{{Marois} et~al.}{2006b}]{Marois:2006tx}
{Marois} C.,  {Lafreni{\`e}re} D.,  {Macintosh} B.,   {Doyon} R.,  2006b, \mn@doi [\apj] {10.1086/505191}, \href {https://ui.adsabs.harvard.edu/abs/2006ApJ...647..612M} {647, 612}

\bibitem[\protect\citeauthoryear{{Marois}, {Macintosh}, {Barman}, {Zuckerman}, {Song}, {Patience}, {Lafreni{\`e}re}  \& {Doyon}}{{Marois} et~al.}{2008}]{Marois:2008tk}
{Marois} C.,  {Macintosh} B.,  {Barman} T.,  {Zuckerman} B.,  {Song} I.,  {Patience} J.,  {Lafreni{\`e}re} D.,   {Doyon} R.,  2008, \mn@doi [Science] {10.1126/science.1166585}, \href {https://ui.adsabs.harvard.edu/abs/2008Sci...322.1348M} {322, 1348}

\bibitem[\protect\citeauthoryear{{Marois}, {Zuckerman}, {Konopacky}, {Macintosh}  \& {Barman}}{{Marois} et~al.}{2010}]{Marois:2010ue}
{Marois} C.,  {Zuckerman} B.,  {Konopacky} Q.~M.,  {Macintosh} B.,   {Barman} T.,  2010, \mn@doi [\nat] {10.1038/nature09684}, \href {https://ui.adsabs.harvard.edu/abs/2010Natur.468.1080M} {468, 1080}

\bibitem[\protect\citeauthoryear{{Mesa} et~al.,}{{Mesa} et~al.}{2019}]{Mesa:2019a}
{Mesa} D.,  et~al., 2019, \mn@doi [\aap] {10.1051/0004-6361/201936764}, \href {https://ui.adsabs.harvard.edu/abs/2019A&A...632A..25M} {632, A25}

\bibitem[\protect\citeauthoryear{{Milli}, {Mouillet}, {Lagrange}, {Boccaletti}, {Mawet}, {Chauvin}  \& {Bonnefoy}}{{Milli} et~al.}{2012}]{Milli:2012ww}
{Milli} J.,  {Mouillet} D.,  {Lagrange} A.~M.,  {Boccaletti} A.,  {Mawet} D.,  {Chauvin} G.,   {Bonnefoy} M.,  2012, \mn@doi [\aap] {10.1051/0004-6361/201219687}, \href {https://ui.adsabs.harvard.edu/abs/2012A&A...545A.111M} {545, A111}

\bibitem[\protect\citeauthoryear{{M{\"u}ller} et~al.,}{{M{\"u}ller} et~al.}{2018}]{Muller:2018wg}
{M{\"u}ller} A.,  et~al., 2018, \mn@doi [\aap] {10.1051/0004-6361/201833584}, \href {https://ui.adsabs.harvard.edu/abs/2018A&A...617L...2M} {617, L2}

\bibitem[\protect\citeauthoryear{{Nelson}, {Robertson}, {Payne}, {Pritchard}, {Deck}, {Ford}, {Wright}  \& {Isaacson}}{{Nelson} et~al.}{2016}]{Nelson:2016}
{Nelson} B.~E.,  {Robertson} P.~M.,  {Payne} M.~J.,  {Pritchard} S.~M.,  {Deck} K.~M.,  {Ford} E.~B.,  {Wright} J.~T.,   {Isaacson} H.~T.,  2016, \mn@doi [\mnras] {10.1093/mnras/stv2367}, \href {https://ui.adsabs.harvard.edu/abs/2016MNRAS.455.2484N} {455, 2484}

\bibitem[\protect\citeauthoryear{{Oberg}, {Kamp}, {Cazaux}  \& {Rab}}{{Oberg} et~al.}{2020}]{Oberg:2020}
{Oberg} N.,  {Kamp} I.,  {Cazaux} S.,   {Rab} C.,  2020, \mn@doi [\aap] {10.1051/0004-6361/202037883}, \href {https://ui.adsabs.harvard.edu/abs/2020A&A...638A.135O} {638, A135}

\bibitem[\protect\citeauthoryear{{Oberg}, {Kamp}, {Cazaux}, {Rab}  \& {Czoske}}{{Oberg} et~al.}{2023}]{Oberg:2023}
{Oberg} N.,  {Kamp} I.,  {Cazaux} S.,  {Rab} C.,   {Czoske} O.,  2023, \mn@doi [\aap] {10.1051/0004-6361/202244845}, \href {https://ui.adsabs.harvard.edu/abs/2023A&A...670A..74O} {670, A74}

\bibitem[\protect\citeauthoryear{{Ogilvie} \& {Lubow}}{{Ogilvie} \& {Lubow}}{2002}]{Ogilvie:2002tq}
{Ogilvie} G.~I.,  {Lubow} S.~H.,  2002, \mn@doi [\mnras] {10.1046/j.1365-8711.2002.05148.x}, \href {https://ui.adsabs.harvard.edu/abs/2002MNRAS.330..950O} {330, 950}

\bibitem[\protect\citeauthoryear{{Pairet}, {Cantalloube}  \& {Jacques}}{{Pairet} et~al.}{2021}]{Pairet:2021}
{Pairet} B.,  {Cantalloube} F.,   {Jacques} L.,  2021, \mn@doi [\mnras] {10.1093/mnras/stab607}, \href {https://ui.adsabs.harvard.edu/abs/2021MNRAS.503.3724P} {503, 3724}

\bibitem[\protect\citeauthoryear{{Penoyre}, {Belokurov}, {Wyn Evans}, {Everall}  \& {Koposov}}{{Penoyre} et~al.}{2020}]{penoyre20}
{Penoyre} Z.,  {Belokurov} V.,  {Wyn Evans} N.,  {Everall} A.,   {Koposov} S.~E.,  2020, \mn@doi [\mnras] {10.1093/mnras/staa1148}, \href {https://ui.adsabs.harvard.edu/abs/2020MNRAS.495..321P} {495, 321}

\bibitem[\protect\citeauthoryear{{Penoyre}, {Belokurov}  \& {Evans}}{{Penoyre} et~al.}{2022}]{penoyre22}
{Penoyre} Z.,  {Belokurov} V.,   {Evans} N.~W.,  2022, \mn@doi [\mnras] {10.1093/mnras/stac1147}, \href {https://ui.adsabs.harvard.edu/abs/2022MNRAS.513.5270P} {513, 5270}

\bibitem[\protect\citeauthoryear{{Perotti} et~al.,}{{Perotti} et~al.}{2023}]{Perotti:2023}
{Perotti} G.,  et~al., 2023, \mn@doi [\nat] {10.1038/s41586-023-06317-9}, \href {https://ui.adsabs.harvard.edu/abs/2023Natur.620..516P} {620, 516}

\bibitem[\protect\citeauthoryear{{Quanz}, {Amara}, {Meyer}, {Girard}, {Kenworthy}  \& {Kasper}}{{Quanz} et~al.}{2015}]{Quanz:2015tu}
{Quanz} S.~P.,  {Amara} A.,  {Meyer} M.~R.,  {Girard} J.~H.,  {Kenworthy} M.~A.,   {Kasper} M.,  2015, \mn@doi [\apj] {10.1088/0004-637X/807/1/64}, \href {https://ui.adsabs.harvard.edu/abs/2015ApJ...807...64Q} {807, 64}

\bibitem[\protect\citeauthoryear{{Ren}}{{Ren}}{2023}]{Ren:2023b}
{Ren} B.~B.,  2023, \mn@doi [\aap] {10.1051/0004-6361/202347354}, \href {https://ui.adsabs.harvard.edu/abs/2023A&A...679A..18R} {679, A18}

\bibitem[\protect\citeauthoryear{{Ren} et~al.,}{{Ren} et~al.}{2023}]{Ren:2023a}
{Ren} B.~B.,  et~al., 2023, \mn@doi [\aap] {10.1051/0004-6361/202347353}, \href {https://ui.adsabs.harvard.edu/abs/2023A&A...680A.114R} {680, A114}

\bibitem[\protect\citeauthoryear{{Samland}, {Brandt}, {Milli}, {Delorme}  \& {Vigan}}{{Samland} et~al.}{2022}]{Samland:2022}
{Samland} M.,  {Brandt} T.~D.,  {Milli} J.,  {Delorme} P.,   {Vigan} A.,  2022, \mn@doi [\aap] {10.1051/0004-6361/202244587}, \href {https://ui.adsabs.harvard.edu/abs/2022A&A...668A..84S} {668, A84}

\bibitem[\protect\citeauthoryear{{Soummer}, {Pueyo}  \& {Larkin}}{{Soummer} et~al.}{2012}]{Soummer:2012wx}
{Soummer} R.,  {Pueyo} L.,   {Larkin} J.,  2012, \mn@doi [\apjl] {10.1088/2041-8205/755/2/L28}, \href {https://ui.adsabs.harvard.edu/abs/2012ApJ...755L..28S} {755, L28}

\bibitem[\protect\citeauthoryear{{Sparks} \& {Ford}}{{Sparks} \& {Ford}}{2002}]{Sparks:2002}
{Sparks} W.~B.,  {Ford} H.~C.,  2002, \mn@doi [\apj] {10.1086/342401}, \href {https://ui.adsabs.harvard.edu/abs/2002ApJ...578..543S} {578, 543}

\bibitem[\protect\citeauthoryear{{Stolker}, {Bonse}, {Quanz}, {Amara}, {Cugno}, {Bohn}  \& {Boehle}}{{Stolker} et~al.}{2019}]{Stolker:2019}
{Stolker} T.,  {Bonse} M.~J.,  {Quanz} S.~P.,  {Amara} A.,  {Cugno} G.,  {Bohn} A.~J.,   {Boehle} A.,  2019, \mn@doi [\aap] {10.1051/0004-6361/201834136}, \href {https://ui.adsabs.harvard.edu/abs/2019A&A...621A..59S} {621, A59}

\bibitem[\protect\citeauthoryear{{Surjanovic}, {Syed}, {Bouchard-C{\^o}t{\'e}}  \& {Campbell}}{{Surjanovic} et~al.}{2022}]{Surjanovic:2022}
{Surjanovic} N.,  {Syed} S.,  {Bouchard-C{\^o}t{\'e}} A.,   {Campbell} T.,  2022, \mn@doi [arXiv e-prints] {10.48550/arXiv.2206.00080}, \href {https://ui.adsabs.harvard.edu/abs/2022arXiv220600080S} {p. arXiv:2206.00080}

\bibitem[\protect\citeauthoryear{{Surjanovic}, {Biron-Lattes}, {Tiede}, {Syed}, {Campbell}  \& {Bouchard-C{\^o}t{\'e}}}{{Surjanovic} et~al.}{2023}]{Surjanovic:2023}
{Surjanovic} N.,  {Biron-Lattes} M.,  {Tiede} P.,  {Syed} S.,  {Campbell} T.,   {Bouchard-C{\^o}t{\'e}} A.,  2023, \mn@doi [arXiv e-prints] {10.48550/arXiv.2308.09769}, \href {https://ui.adsabs.harvard.edu/abs/2023arXiv230809769S} {p. arXiv:2308.09769}

\bibitem[\protect\citeauthoryear{Syed, Bouchard-Côté, Deligiannidis  \& Doucet}{Syed et~al.}{2021}]{10.1111/rssb.12464}
Syed S.,  Bouchard-Côté A.,  Deligiannidis G.,   Doucet A.,  2021, \mn@doi [Journal of the Royal Statistical Society Series B: Statistical Methodology] {10.1111/rssb.12464}, 84, 321

\bibitem[\protect\citeauthoryear{{Szul{\'a}gyi}, {Dullemond}, {Pohl}  \& {Quanz}}{{Szul{\'a}gyi} et~al.}{2019}]{Szulagyi:2019vo}
{Szul{\'a}gyi} J.,  {Dullemond} C.~P.,  {Pohl} A.,   {Quanz} S.~P.,  2019, \mn@doi [\mnras] {10.1093/mnras/stz1326}, \href {https://ui.adsabs.harvard.edu/abs/2019MNRAS.487.1248S} {487, 1248}

\bibitem[\protect\citeauthoryear{{Thompson} et~al.,}{{Thompson} et~al.}{2023}]{Thompson:2023a}
{Thompson} W.,  et~al., 2023, \mn@doi [\aj] {10.3847/1538-3881/acf5cc}, \href {https://ui.adsabs.harvard.edu/abs/2023AJ....166..164T} {166, 164}

\bibitem[\protect\citeauthoryear{{Toci}, {Lodato}, {Christiaens}, {Fedele}, {Pinte}, {Price}  \& {Testi}}{{Toci} et~al.}{2020}]{Toci:2020wu}
{Toci} C.,  {Lodato} G.,  {Christiaens} V.,  {Fedele} D.,  {Pinte} C.,  {Price} D.~J.,   {Testi} L.,  2020, \mn@doi [\mnras] {10.1093/mnras/staa2933}, \href {https://ui.adsabs.harvard.edu/abs/2020MNRAS.499.2015T} {499, 2015}

\bibitem[\protect\citeauthoryear{{Wahhaj} et~al.,}{{Wahhaj} et~al.}{2021}]{Wahhaj:2021}
{Wahhaj} Z.,  et~al., 2021, \mn@doi [\aap] {10.1051/0004-6361/202038794}, \href {https://ui.adsabs.harvard.edu/abs/2021A&A...648A..26W} {648, A26}

\bibitem[\protect\citeauthoryear{{Wahhaj} et~al.,}{{Wahhaj} et~al.}{2024}]{Wahhaj:2024}
{Wahhaj} Z.,  et~al., 2024, \mn@doi [\aap] {10.1051/0004-6361/202349018}, \href {https://ui.adsabs.harvard.edu/abs/2024A&A...687A.257W} {687, A257}

\bibitem[\protect\citeauthoryear{{Wang} et~al.,}{{Wang} et~al.}{2020}]{Wang:2020}
{Wang} J.~J.,  et~al., 2020, \mn@doi [\aj] {10.3847/1538-3881/ab8aef}, \href {https://ui.adsabs.harvard.edu/abs/2020AJ....159..263W} {159, 263}

\bibitem[\protect\citeauthoryear{{Wang} et~al.,}{{Wang} et~al.}{2021}]{Wang:2021a}
{Wang} J.~J.,  et~al., 2021, \mn@doi [\aj] {10.3847/1538-3881/abdb2d}, \href {https://ui.adsabs.harvard.edu/abs/2021AJ....161..148W} {161, 148}

\bibitem[\protect\citeauthoryear{{Weiss}, {Millholland}, {Petigura}, {Adams}, {Batygin}, {Block}  \& {Mordasini}}{{Weiss} et~al.}{2023}]{Weiss:2023}
{Weiss} L.~M.,  {Millholland} S.~C.,  {Petigura} E.~A.,  {Adams} F.~C.,  {Batygin} K.,  {Block} A.~M.,   {Mordasini} C.,  2023, in {Inutsuka} S.,  {Aikawa} Y.,  {Muto} T.,  {Tomida} K.,   {Tamura} M.,  eds,  Astronomical Society of the Pacific Conference Series Vol. 534, Protostars and Planets VII. p.~863

\bibitem[\protect\citeauthoryear{{Wertz}, {Absil}, {G{\'o}mez Gonz{\'a}lez}, {Milli}, {Girard}, {Mawet}  \& {Pueyo}}{{Wertz} et~al.}{2017}]{2017A&A...598A..83W}
{Wertz} O.,  {Absil} O.,  {G{\'o}mez Gonz{\'a}lez} C.~A.,  {Milli} J.,  {Girard} J.~H.,  {Mawet} D.,   {Pueyo} L.,  2017, \mn@doi [\aap] {10.1051/0004-6361/201628730}, \href {https://ui.adsabs.harvard.edu/abs/2017A&A...598A..83W} {598, A83}

\bibitem[\protect\citeauthoryear{{Xie} et~al.,}{{Xie} et~al.}{2022}]{Xie:2022}
{Xie} C.,  et~al., 2022, \mn@doi [\aap] {10.1051/0004-6361/202243379}, \href {https://ui.adsabs.harvard.edu/abs/2022A&A...666A..32X} {666, A32}

\bibitem[\protect\citeauthoryear{{Zhou} et~al.,}{{Zhou} et~al.}{2025}]{Zhou:2025}
{Zhou} Y.,  et~al., 2025, \mn@doi [\apjl] {10.3847/2041-8213/adb134}, \href {https://ui.adsabs.harvard.edu/abs/2025ApJ...980L..39Z} {980, L39}

\makeatother
\end{thebibliography}



\appendix

\section{Gemini/NICI L'-band angular differential imaging observations}\label{sec:nici_data}
We re-processed non-coronagraphic Gemini/NICI observations from the night of 31 March 2012 to extend our time baseline to almost ten years. These \textit{L'}-data, in pupil stabilised mode, were published in \citet{Hashimoto:2012wi}. Details on the pre-processing are in \citet{Keppler:2018vt}. The final cube consisted of 107 frames (comprised of 25 co-adds of 0.76s) with 99\fdg4 of parallatic angle variation after bad frame rejection. We used the same approach to post-processing as with NaCo, using annular PCA-ADI in a 1\farcs2 wide annular centred on the star and a rotation threshold of 0.5$\times$FWHM at the approximate separation of candidate~d and explored $n_{\rm pc}$ = 1--50 with $n_{\rm pc}$ = 10 shown in Fig. \ref{fig:nici_appendix}. A point-like feature is identified at $\sim$ 336$\fdg$4 $\pm$ 10$\fdg$87 with a radial separation of 146.99 mas $\pm$ 19.91 mas. We suspect this feature to trace a residual speckle rather than thermal emission from d, due to its comparatively wider separation that overlaps with the predicted orbital solutions for b. The feature is also of low significance given other brighter signals at the same separation. 

\begin{figure}\includegraphics[width=0.9\columnwidth]{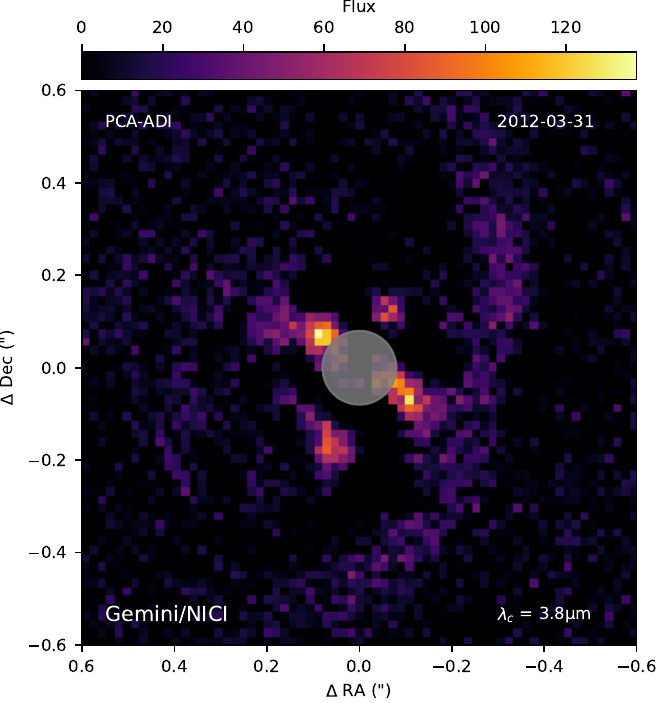}
    \caption{Our post-processed image of the Gemini/NICI data from 2012 Mar 31, displayed the same as all other epochs in Fig. \ref{fig:observations}. Several signals are retrieved at the same approximate separation as PDS~70~d.}
    \label{fig:nici_appendix}
\end{figure}

\section{ANDROMEDA and Mayonnaise reduction of SPHERE data}\label{sec:andromeda+mayo_data}

\begin{figure*}\includegraphics[width=0.9\textwidth]{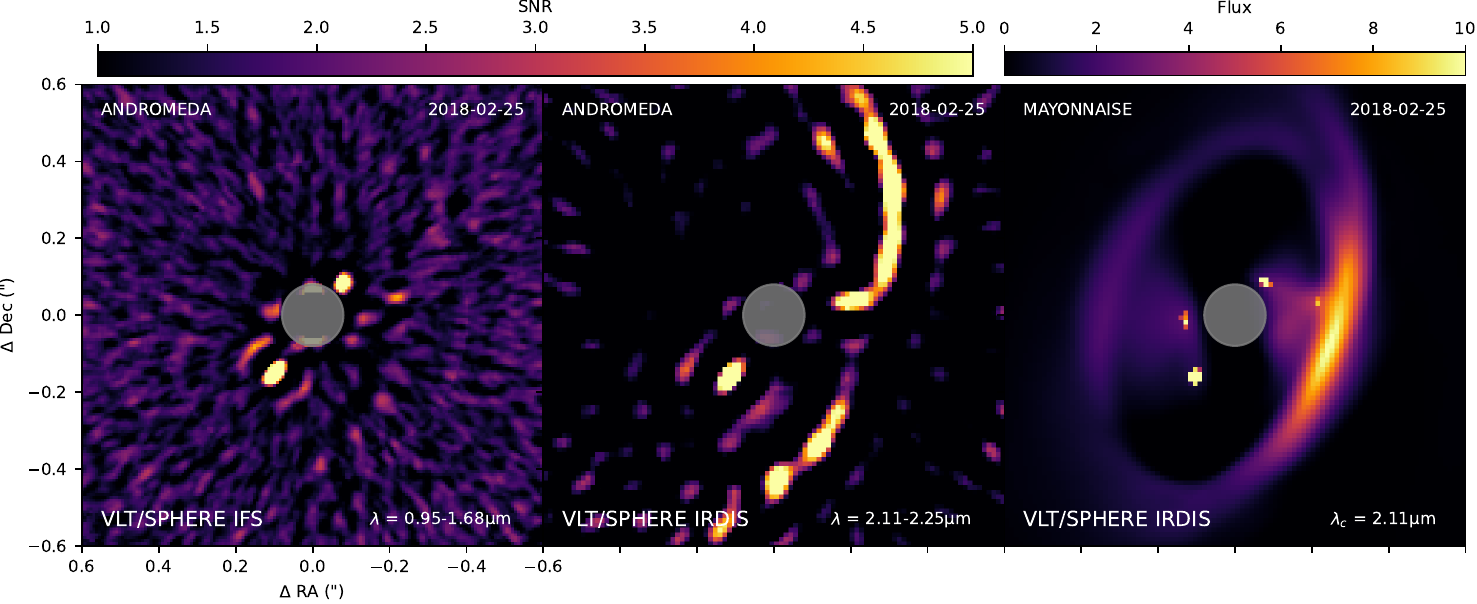}
    \caption{\textit{Left} and \textit{Middle:} IFS and IRDIS (\emph{K$_1$K$_2$}) SNR maps obtained with the \textsc{andromeda} algorithm. \textit{Right}: \textsc{mayonnaise} \citep{Pairet:2021} identifies PDS~70~d as a point-source and not a disc feature, but it does the same for the emission in the east of the star.}
    \label{fig:andromeda+mago_fig}
\end{figure*}

The SNR maps obtained with \textsc{andromeda} for the 2018 Feb 25 SPHERE data are shown in Fig.~\ref{fig:andromeda+mago_fig}. The two known protoplanets are detected by each subinstrument, whereas signal from candidate d is significantly more robust in the IFS wavelength range. The successful detection indicates that d traces a rotating point-source. We also tested the \textsc{mayonnaise} algorithm \citep{Pairet:2021} on IRDIS \emph{K$_1$} data to disentangle point-sources from extended signals. Figure~\ref{fig:andromeda+mago_fig} (right) shows the results of this test. The algorithm does identify d as well as b \& c as non-disc features, but also does the same for emission to the east of the star \citep[PLF~2,][]{Pairet:2021}.

\section{RUWE predictions}
\label{sec:ruwe}
As the candidate is closer to the star than the existing planets, we tested whether a third protoplanet is compatible with the re-normalized unit weight error (RUWE) of 1.35 from \textit{Gaia} DR3 \citep{Gaia-Collaboration:2023}. This value is calculated as a $\chi^2$ error by comparing the position of the photocentre to a single star solution \citep{penoyre20} and re-normalised for a star's brightness and colour \citep{2018RenormalisingTA}. A RUWE value of near 1 indicates the photocentre's position can be explained by a single star track.  Values greater than $\sim$1.25 could be indicative of a companion \citep{penoyre22}. 

In order to understand the effect of a planetary mass companion on RUWE, we simulated the value of RUWE of PDS~70~A as a function of planet mass and period, assuming a single planet model and a distance of 113~pc. Inclination was fixed at 130$^{\circ}$ and we ignore the presence of the disc, assuming RUWE is solely influenced by a companion.  This is presented in Fig.~\ref{fig:ruwe}.  This result demonstrates that the existing protoplanets, even at their highest mass estimates, have effectively no impact on the RUWE due to their separation from the star. Adding another protoplanet at 13~au does not significantly affect the predicted RUWE. The distance to the source, and other factors such as variability \citep{Gaidos:2024} and obstructing disc material, are likely dominating the RUWE budget more than any of the proposed or confirmed planets.

\begin{figure}\label{fig:ruwe}
    \includegraphics[width=0.9\columnwidth]{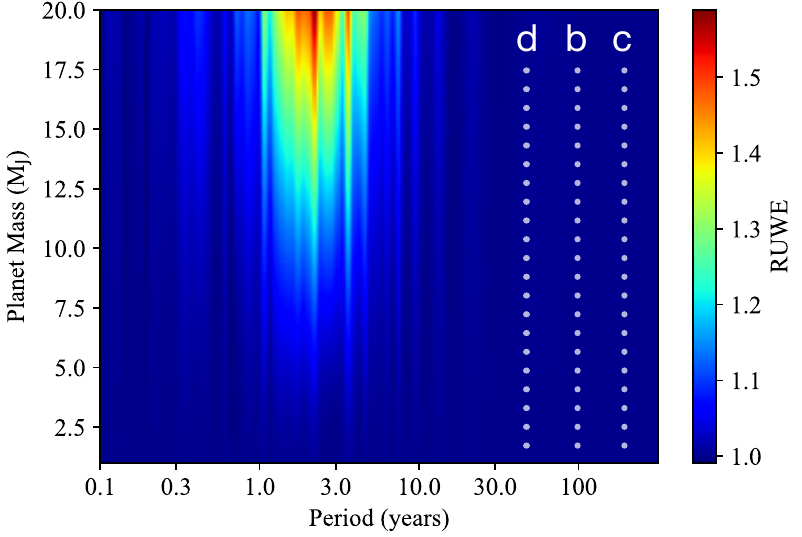}
    \caption{RUWE of PDS~70~A as a function of planet mass and period. This demonstrates that the detected planets, or PDS~70~d, cannot alone account for the measured RUWE of 1.35.}
\end{figure}

\section{Octofitter Corner Plot}\label{sec:corner_plot}
In Fig. \ref{fig:corner_plot} we show the marginal posterior distributions of the eccentricity, mutual inclination with the disc and and orbital period of PDS 70 b, PDS 70 c and PDS 70 d, from our model described in Section \ref{sec:astrometry}.

\begin{figure}\includegraphics[width=\columnwidth]{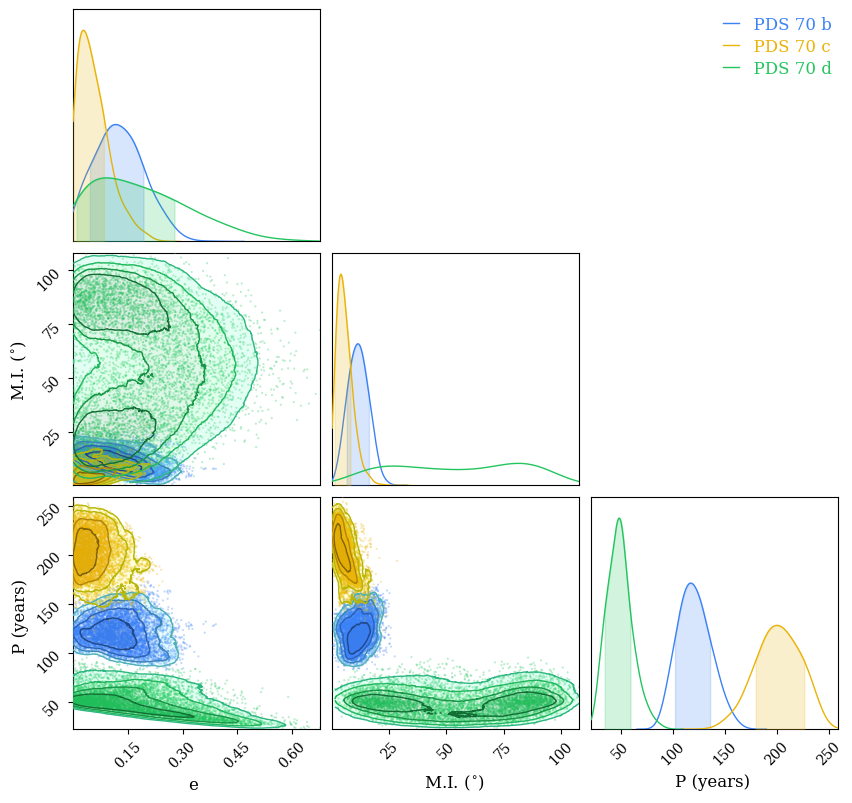}
    \caption{Comparison between the posterior samples of the eccentricity, mutual inclination with the disc, assuming $i = 128\fdg3$ and $\Omega = 156\fdg7$ for the disc \citep{2019A&A...625A.118K}, and orbital period of the protoplanets b, c and candidate d, from our \textsc{Octofitter} results, described in Section \ref{sec:astrometry}.
    }
    \label{fig:corner_plot}
\end{figure}

\section{Individual IFS Channels}\label{sec:allchans}
Figure \ref{fig:all_channels} shows all 39 SPHERE/IFS channels from the 2022 Feb 28 epoch after subtraction of the reference star. Candidate~d is detected in channels that do not suffer from telluric absorption, adding support to it being a genuine feature rather than an observational artefact. Figure \ref{fig:sub_all_channels} shows the same cube with the best estimate for the flux of d subtracted. 

\begin{figure*}
    \includegraphics[width=0.95\textwidth]{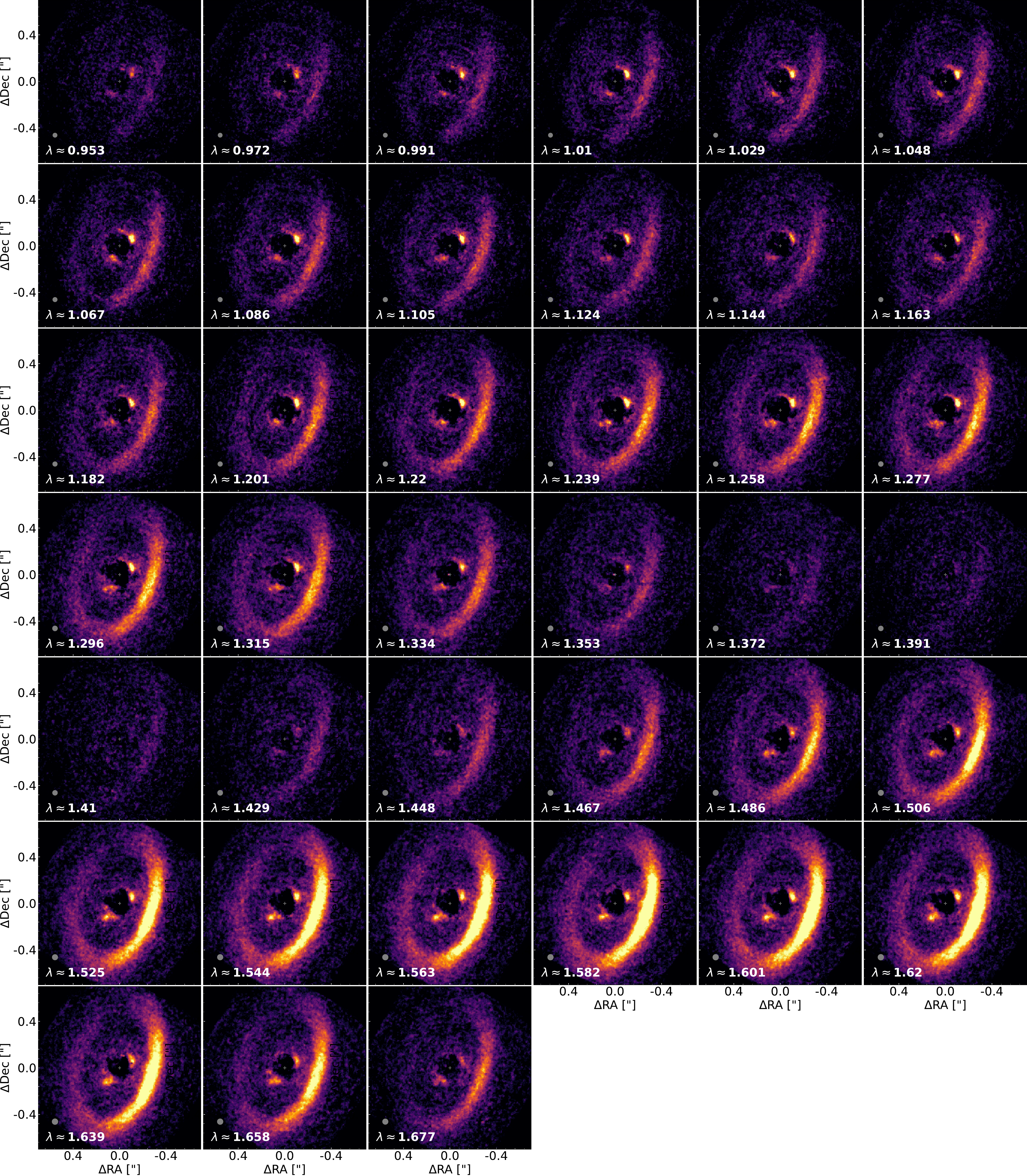}
    \caption{Iterative PCA-RDI for the 2022 Feb 28 epoch in Fig. \ref{fig:observations} truncated to seven principal components and obtained with three iterations, for each individual spectral channel. For all frames, the PSF of the reference star UCAC2~14412811 observed during the sequence has been subtracted. Candidate~d is visible in all channels that are not contaminated by telluric absorption, whereas PDS~70~b is conspicuous at longer wavelengths (i.e., \textit{H}-band). The grey patch represents the FWHM of the non-coronagraphic PSF and is unique for each channel. No numerical mask has been used to represent the location of the coronagraphic mask. Pixel intensities are on a linear scale with an intensity cut from the 0–99.5 percentiles of the entire cube. Figure made using \textsc{hciplot} (\url{https://github.com/carlos-gg/hciplot}).}
    \label{fig:all_channels}
\end{figure*}

\begin{figure*}
    \includegraphics[width=0.95\textwidth]{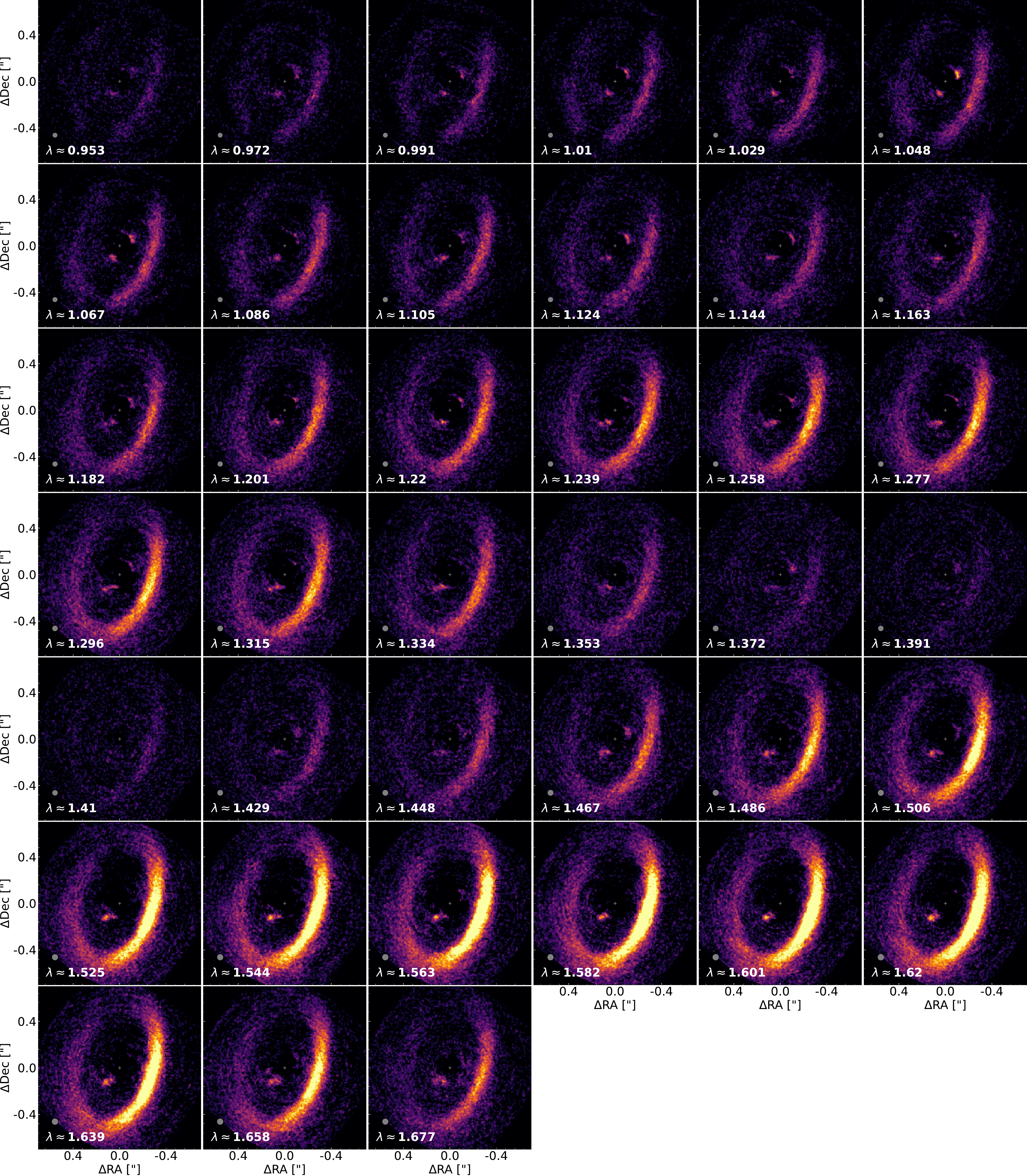}
    \caption{Same as Fig. \ref{fig:all_channels}, but with our best estimate for the flux of PDS~70~d from Section \ref{sec:spectrum} for each channel subtracted. The flux is generally well subtracted with the exception of some extended emission in the vicinity of d at short wavelengths.}
    \label{fig:sub_all_channels}
\end{figure*}



\bsp	
\label{lastpage}
\end{document}